\documentclass[aps,pra,twocolumn,superscriptaddress,footnotebib]{revtex4-1}
\usepackage{amsmath}
\usepackage{amssymb}
\usepackage{amsfonts}
\usepackage{bm}
\usepackage{color}
\usepackage{graphicx}
\usepackage{epstopdf}
\usepackage{epsfig}
\usepackage[naturalnames]{hyperref}
\usepackage{soul}
\usepackage{hypcap}
\usepackage{verbatim}
\usepackage{tabularx}
\usepackage{bbm}
\usepackage{esvect}
\usepackage{subfigure}
\usepackage{braket}
\graphicspath{{Figures/}}

\newcommand{\CE}[0]{\mathcal{E}}
\newcommand{\avg}[0]{\mathbb{E}}
\newcommand{\tr}[0]{\text{Tr}}

\newcommand{\adj}[1]{{#1}^{\dagger}}
\newcommand{\magn}[1]{\left| #1 \right|}

\def\bea{\begin{eqnarray}}
\def\eea{\end{eqnarray}}
\def\nn{\nonumber}
\def\ba{\begin{array}}
\def\ea{\end{array}}
\def\nn{\nonumber}
\def\Tr{\text{Tr}}

\def\sgn{\text{sgn}}

\def\id{\mathbb{I}}

\hypersetup{colorlinks=true, citecolor=blue, urlcolor=blue, linkcolor=blue}
\begin{document}

\title{Linear Growth of Circuit Complexity from Brownian Dynamics}

\author{Shao-Kai Jian}
\author{Gregory Bentsen}
\author{Brian Swingle}
\affiliation{Department of Physics, Brandeis University, Waltham, Massachusetts 02453, USA}

\begin{abstract}
\end{abstract}

\maketitle

{\bf We calculate the frame potential for Brownian clusters of $N$ spins or fermions with time-dependent all-to-all interactions. In both cases the problem can be mapped to an effective statistical mechanics problem which we study using a path integral approach. We argue that the $k$th frame potential comes within $\epsilon$ of the Haar value after a time of order $t \sim k N + k \log k + \log \epsilon^{-1}$. Using a bound on the diamond norm, this implies that such circuits are capable of coming very close to a unitary $k$-design after a time of order $t \sim k N$. We also consider the same question for systems with a time-independent Hamiltonian and argue that a small amount of time-dependent randomness is sufficient to generate a $k$-design in linear time provided the underlying Hamiltonian is quantum chaotic. These models provide explicit examples of linear complexity growth that are also analytically tractable.}


\section{Introduction}
\label{sec:intro}

The ability to sample from the Haar distribution of unitaries on a Hilbert space $\mathcal{H}$ is a widely useful capability \cite{bernstein1997quantum,poulin2011quantum,aaronson2016complexity} and is often a crucial ingredient in modern quantum information processing tasks including randomized benchmarking \cite{emerson2005scalable,knill2008randomized,magesan2012characterizing} and demonstrations of quantum advantage \cite{boixo2018characterizing} in noisy intermediate-scale quantum processing devices. More generally, Haar-random unitaries and related matrix distributions are foundational to many areas of modern quantum information science, where they play central roles in our understanding of quantum information scrambling \cite{hayden2007black,dupuis2014one}, quantum chaos in many-body thermalizing systems \cite{roberts2017chaos,cotler2017chaos,brown2018second,liu2018entanglement}, and models of black hole dynamics in holographic quantum gravity \cite{brown2016complexity,brown2016holographic,susskind2016computational,stanford2014complexity,susskind2018black}. However, it is known that for $N$ qubits, the number of elementary $2$-qubit gates needed to generate samples from the Haar distribution is exponential in $N$ \cite{knill1995approximation}. In such a setting, it is important to understand under what conditions one can approximately sample from the Haar distribution with more modest resources \cite{roberts2017chaos,hunter2019unitary,brandao2021models}. For many purposes it is sufficient to reproduce moments of the Haar distribution involving at most $k$ copies of $U$ and $k$ copies of $U^\dagger$, and any ensemble which does so is called a $k$-design \cite{dankert2005efficient,dankert2009exact,gross2007evenly,ambainis2007quantum}. Examples include the calculation of average purity ($k=2$), averages of out-of-time-order correlators ($k=2$), and higher Renyi entropies ($k =$ Renyi index). 

The purpose of this paper is to show that a wide variety of Brownian quantum many-body systems form good approximate $k$-designs in time linear in $k$. This scaling of the time-to-design is essentially optimal in its $k$-dependence up to a $\log k$ factor~\cite{brown2018second, susskind2018black}, so our results establish that these quantum chaotic model systems are optimal generators of quantum randomness \cite{kdesigncomplexity}. Our calculations build on a growing body of work featuring random circuits and Brownian models \cite{harrow2009random,harrow2009efficient,brown2012scrambling,lashkari2013towards,brown2015decoupling,brandao2016local,onorati2017mixing,nakata2017efficient,harrow2018approximate,nahum2018operator,zhou2019emergent,hunter2019unitary,bentsen2021measurement,haferkamp2022random,haferkamp2022linear}, which have been used to establish polynomial complexity growth in specific circuit constructions. Our Brownian spin and fermion models are based on similar technical methods, but the simple mean-field nature of our models allows us to straightforwardly establish linear complexity growth in a large-$N$ limit. These results may have practical relevance in any application calling for $k$-designs \cite{dankert2005efficient,gross2007evenly}. They are also interesting from the perspective of formal computational complexity theory. In particular, by virtue of approximating a $k$-design in linear time, our models also have linearly-growing complexity as measured by more robust information-theoretic notions of strong circuit complexity~\cite{brandao2021models,haferkamp2022linear}.

To characterize the growth of complexity in these systems, we focus on calculating the Frame Potential (FP) \cite{gross2007evenly,scott2008optimizing}, which has recently gained interest as a diagnostic of quantum chaos \cite{roberts2017chaos,cotler2017chaos,hunter2019unitary,brandao2021models}. Qualitatively, the FP provides a measure of distance between a particular channel of interest and the Haar-random ensemble. Specifically, given any unitary channel $\mathcal{E}$, the $k$th Frame Potential $F^{(k)}_{\mathcal{E}}$ measures the 2-norm distance between the channel and the Haar-random ensemble, with minimal value $F^{(k)}_{\mathcal{E}} \geq F^{(k)}_{\mathrm{Haar}} = k!$ if the channel $\mathcal{E}$ is indistinguishable from the Haar-random ensemble. Quantitatively, the difference $F^{(k)}_{\mathcal{E}} - F^{(k)}_{\mathrm{Haar}}$ bounds the diamond norm between the two channels, so any circuit that achieves a near-minimal value $F^{(k)}_{\mathcal{E}} \sim k!$ is strictly indistinguishable from the Haar-random ensemble in precise information-theoretic terms that we review below \cite{hunter2019unitary,brandao2021models}.

\begin{figure*}
    \centering
    \includegraphics[width=0.7\textwidth]{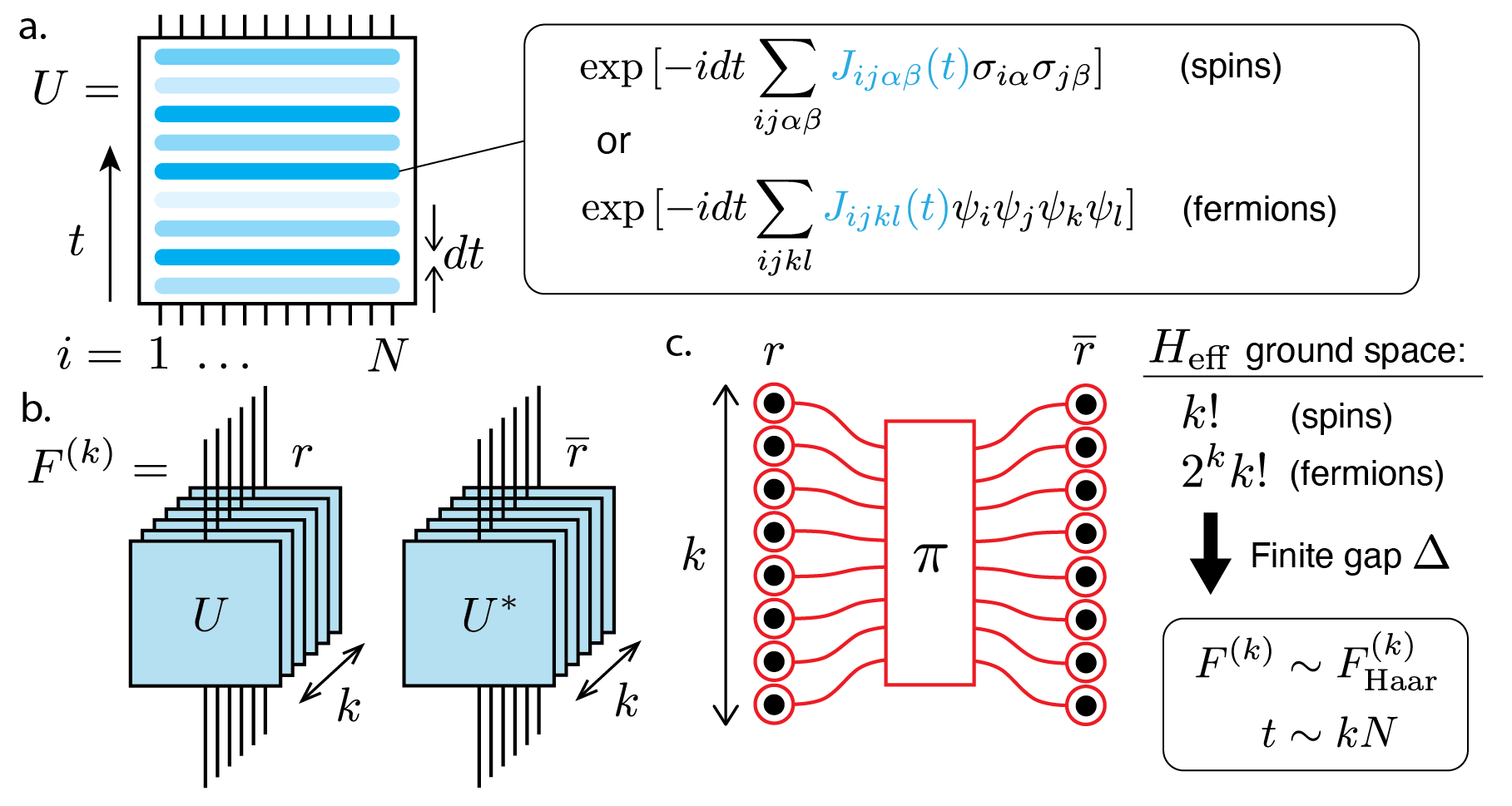}
    \caption{Linear growth of complexity in Brownian spin and fermion clusters. The cluster evolves under time-dependent unitary dynamics $U$ (a) composed of Brownian 2-body interactions (spins) or Brownian 4-body interactions (fermions). To compute the Frame Potential $F^{(k)}$ we consider $2k$ replicas (b) labelled by $r,\overline{r} = 1,\ldots,k$ for forward and backward time evolution, respectively. Following disorder averaging, the $2k$-replica system is governed by an effective Hamiltonian $H_{\mathrm{eff}}$ whose ground state manifold (c) is $k!$-fold degenerate for spins and $2^k k!$-fold degenerate for fermions. For spins, each ground state corresponds to one of $k!$ possible pairings (red) between the $r,\overline{r}$ replicas, labelled by elements $\pi$ of the symmetric group $S_k$. For fermions there is an extra factor of $2^k$ coming from $\pm$ signs attached to each pairing. In both cases, a finite gap $\Delta$ to the excited state manifold guarantees that the clusters form $k$-designs in a time $t \sim k N$ linear in both $k$ and $N$ given $k \le e^N$.}
    \label{fig:fig1}
\end{figure*}

In this paper, we analyze in detail the Frame Potential for Brownian spin and fermion models with all-to-all interactions. In both cases we arrive at effective statistical models describing the FP similar in spirit to the mapping in Refs.~\cite{hunter2019unitary, banchi2017driven}, except that the all-to-all and Brownian character of our models leads to much simpler statistical models that are analytically tractable. In particular, we show that the $k$th Frame Potential is equivalent to the partition function $Z = \tr{\left[ e^{-\beta H_{\mathrm{eff}}} \right]}$ of an effective Hamiltonian with a tractable mean-field structure, where the inverse temperature $\beta = 2 t$ is given by the depth $t$ of the Brownian circuit.
As a result, we can make precise statements about the FP -- and therefore the system's circuit complexity -- just by studying the spectrum of the effective Hamiltonian $H_{\mathrm{eff}}$. In particular, at long times $t \rightarrow \infty$ the FP is governed entirely by the Hamiltonian's ground state manifold and gap $\Delta$ to the excited state manifold. We also show how the resulting partition function can be accurately calculated using a path integral approach. 

We also discuss straightforward extensions of our construction to time-independent Hamiltonians \cite{nakata2017efficient,cotler2017chaos,brandao2021models}. While a single fixed Hamiltonian or even an ensemble of Hamiltonians cannot form a good approximate $k$-design at long time \cite{roberts2017chaos, cotler2017chaos}, we argue here that time-independent Hamiltonians perturbed by a small amount of Brownian noise can form good approximate $k$-designs in time linear in $k$. For simplicity we focus here on an ensemble of Hamiltonians chosen from the Gaussian random matrix ensemble, but we expect similar arguments to hold for generic strongly chaotic quantum Hamiltonians.


Before analyzing our Brownian models, we first review relevant formal definitions of complexity and summarize our main technical results in Section \ref{sec:defns}. We then turn to specific models, considering first a Brownian model on $N$ spins in Section \ref{sec:brownianspins} for which we obtain the effective Hamiltonian controlling the FP. A parallel path integral calculation reproduces the partition function for this effective Hamiltonian, which is the FP. We then consider an analogous Brownian model on $N$ fermions in Section \ref{sec:brownianfermions} and again compute the FP via a path integral method. The main difference is an extra fermion parity symmetry arising from the fermionic degrees of freedom. We also analyze $1/N$ corrections in the fermionic path integral and check that they do not modify our results. Finally, in Section \ref{sec:timeindepham} we turn to a discussion of time-independent Hamiltonians perturbed by Brownian noise, focusing on the simple toy model of a Gaussian random Hamiltonian. We conclude in Section \ref{sec:discussion} with a summary of our results and a discussion of future directions and open problems.

\section{Formal Definitions of $k$-Designs}
\label{sec:defns}

Before going further, we need to define good approximate $k$-designs. Given an ensemble $\mathcal{E}$ of unitaries acting on $\mathcal{H}$, the $k$-th \emph{moment map} $\tilde{M}_{\mathcal{E},k}$ acts on an operator $X$ on $\mathcal{H}^{\otimes k}$ as \cite{haferkamp2022random}
\begin{equation} \label{eq:k-channel}
    \tilde{M}_{\mathcal{E},k}(X) = \mathbb{E}_{U\in \mathcal{E}}\left[U^{\otimes k} X (U^\dagger)^{\otimes k} \right].
\end{equation}
If each $U$ and $U^\dagger$ are equiprobable in $\mathcal{E}$, then the moment map is a Hermitian superoperator with a real spectrum and a complete set of eigenoperators. An ensemble is a $k$-design if $\tilde M_{\mathcal{E},k}=\tilde M_{\text{Haar},k}$, and it is an approximate $k$-design if
\begin{equation}
    \| \tilde{M}_{\mathcal{E},k} - \tilde{M}_{\text{Haar},k} \|_{\diamond} < \epsilon_{\diamond},
\end{equation}
where $\|\cdot \|_{\diamond}$ denotes the diamond norm \cite{aharonov1998quantum,watrous2018theory}. This norm bounds the single shot distinguishability of two channels even in the presence of ancilla. Another weaker notion of approximate $k$-design is an approximate $k$-copy tensor product expander, which requires 
\begin{equation}
    \| \tilde{M}_{\mathcal{E},k} - \tilde{M}_{\text{Haar},k} \|_{\infty} < \epsilon_{\infty},
\end{equation}
where $\| \cdot \|_{\infty}$ is the operator norm on channels.

Here we study these notions in the context of two all-to-all Brownian models, one built from spins \cite{bentsen2021measurement} and one built from fermions \cite{saad2018semiclassical}. 
In both cases, we show that time evolution for a time $t$ satisfying
\begin{equation}\label{eq:time}
    t > t_0 ( c_1 k N + c_2 \log \epsilon_{\diamond}^{-1} )
\end{equation}
is sufficient to generate an approximate $k$-design for any $k \le D$. Here $c_{1,2}$ are numbers, $D$ is the dimension of Hilbert space and $N$ is the number of spins or Majorana fermions. 
For the weaker notion of $k$-copy tensor product expander, the time is less than~(\ref{eq:time}) but still linear in $k N$. 
We also analyze ensembles generated by perturbing chaotic time-independent Hamiltonians with a small amount of random noise, and we find that various notions of quantum chaos are sufficient to guarantee that the resulting ensemble forms an approximate $k$-design in time linear in $k$.

The notion of a frame potential (FP) plays a central role in our analysis. Given an ensemble $\mathcal{E}$, the $k$-th FP is
\begin{equation}
    F^{(k)}_\CE = \avg_{U \in \CE,V \in \CE} |\tr(V^\dagger U)|^{2k}.
\end{equation}
If $\CE$ is the Haar ensemble, then $F^{(k)}_{\text{Haar}} = k!$, which is the minimal possible value~\cite{roberts2017chaos}. An equivalent condition to $\CE$ being a $k$-design is $F^{(k)}_\CE=k!$, the Haar value. 
Moreover, if an ensemble has a frame potential that is sufficiently close to $k!$, then it is necessarily a good approximate $k$-design in the above diamond norm sense. This is because the FP is related to a $2$-norm, and the two norms can be related as 
\begin{equation} \label{eq:diamond-bound}
    ||\tilde{M}_{\CE,k} - \tilde{M}_{\text{Haar},k} ||^2_{\diamond} \leq D^{2k} \left[F^{(k)}_\CE - F^{(k)}_\text{Haar}\right],
\end{equation}
where $D$ is the dimension of the Hilbert space~\cite{hunter2019unitary}.


\section{Brownian Spin Cluster}
\label{sec:brownianspins}

Having established a concrete definition for complexity via $k$-designs and the Frame Potential, we now turn to specific models where the FP can be computed explicitly. We first consider a Brownian $2$-body spin cluster, which we analyze using an effective Hamiltonian and a path integral approach.

\subsection{Effective Hamiltonian approach} \label{sec:spin-hamiltonian}

Consider a system of $N$ spins with a time-dependent Hamiltonian consisting of a sum over random $2$-body terms,
\begin{equation} \label{eq:spin-model}
    H(t) =  
    \sum_{i<j,\alpha\beta}J_{ij\alpha\beta}(t) \bm\sigma_{i\alpha}\bm\sigma_{j\beta},
\end{equation}
where $i,j=1,...,N$ denote the spins, and $\bm\sigma_{i\alpha} $, $\alpha= 1,2,3$, denotes $\alpha$-th Pauli operator. $J_{ij\alpha\beta}$ is a Brownian Gaussian variable with mean zero and variance
\bea \label{eq:spin-J}
    \avg [J_{ij\alpha\beta}(t) J_{i'j'\alpha'\beta'}(0)] = \delta_{ii'}\delta_{jj'} \delta_{\alpha\alpha'} \delta_{\beta\beta'} \delta(t) \frac{J}{N}.
\eea
Here $\avg$ denotes average over the couplings $J_{ij\alpha\beta}(t)$.

Because the couplings are white noise correlated with zero average, in any average of time evolution operators we can always reinterpret $\tr(V^\dagger(t)U(t)) = \tr(U(2t))$, where $U(2t)$ is a Brownian evolution for twice as long with the couplings in $(0,t)$ determined by the original $U$ and the couplings in $(t,2t)$ determined by the original $V$. The FP is thus
\begin{equation}
    F^{(k)}_{\text{b-spin}} = \avg \left\{\tr[ U(2t) ^{\otimes k} \otimes U(2t)^{*\otimes k} ]\right\},
\end{equation}

Now for the Brownian spin model (\ref{eq:spin-model}), the FP is $F_{\text{b-spin}}^{(k)} = \tr(e^{-2t H_k})$, where
\bea
    H_k &=& 
    \frac{J}{2N} \sum_{i<j,\alpha\beta} \left(\sum_{r=1}^k \{\bm \sigma^r_{i\alpha} \bm \sigma^r_{j\beta} -(\bm\sigma^{\bar{r}}_{i\alpha} \bm\sigma^{\bar{r}}_{j\beta})^*\} \right)^2.
\eea
Here, we have used the fact that $J_{ij\alpha\beta}$ is a Brownian variable~(\ref{eq:spin-J}). This defines our effective statistical model: the frame potential is simply the thermal partition function of the Hamiltonian $H_k$ at inverse temperature $\beta = 2t$.

At first inspection, this Hamiltonian has $k!$ zero energy states given by all possible pairings of forward contours with backward contours into infinite temperature thermofield double states (or EPR state), $\bigotimes_{r \bar s}|\infty\rangle_{r\bar s}$, where $r$ and $\bar s$ appears once in the tensor product, with the following property,
\bea
    \bm \sigma^r_{i\alpha} | \infty \rangle_{r\bar s} = (\bm\sigma^{\bar s}_{i\alpha})^{\ast} |\infty \rangle_{r\bar s},~~ \forall i=1,...,N, \alpha=1,2,3.
\eea
However, these states are not all exactly orthogonal, with overlaps suppressed by powers of $D$. It is known that the dimension of the space of all such pairings is exactly $k!$ provided $k<D+1$~\cite{baik2001algebraic,collins2003moments}. For simplicity, we restrict to $k<D+1$ in the subsequent, but we expect that Brownian models can still form good approximate $k$-designs for larger values of $k$. 

What about excited states? If $\Delta$ is the gap to the first excited state (which might depend on $N$ and $k$), then we can conclude that the ensemble of time evolutions forms an approximate $k$-design once $2 \Delta t > 4k \log D + 2 \log \epsilon_{\diamond}^{-1}$. This is based on a crude estimate of the FP,
\begin{equation}
    \label{eq:framepotbspin}
    F^{(k)}_\text{b-spin} \leq k! + (D^{2k}-k!) e^{-2 \Delta t},
\end{equation} 
combined with the bound \eqref{eq:diamond-bound}. Note that if we have a better estimate for the FP, we can get a better bound on the time to come close to a $k$-design. We now analyze the spectrum of $H_k$ in detail and show that the gap is $\Delta = 12J + O(1/N)$ independent of $k$ and approximately independent of $N$ at large $N$.

\subsubsection{$k=1$}

The effective Hamiltonian with $k=1$ is
\begin{equation}
    H_1 = 
    \frac{J}{N} \sum_{i<j,\alpha\beta} (1 - \bm \sigma^1_{i\alpha} \bm\sigma^1_{j\beta} (\bm \sigma^{\bar{1}}_{i\alpha} \bm\sigma^{\bar{1}}_{j\beta})^*).
\end{equation}
The unique ground state is $|\infty\rangle = |\infty \rangle_{1\bar 1}$. We can generate excited states by acting with a Pauli string $P$ on replica $1$. Running over all possible Pauli strings gives a complete basis for the $1\bar{1}$ Hilbert space. 

Given a Pauli string $P$, each term in $H_1$ either commutes with $P$ or gets flipped,
\begin{equation}
    P(1- O^1 (O^{\bar{1}})^*)P = (1+O^1 (O^{\bar{1}})^*).
\end{equation}
Whereas the original terms annihilate $|\infty\rangle$, the flipped terms have $|\infty\rangle$ as an eigenstate with positive eigenvalue,
\begin{equation}
    (1+O^1 (O^{\bar{1}})^*) |\infty\rangle = 2 |\infty\rangle.
\end{equation}
Hence, for every term that gets flipped, the energy goes up by $2J/N$. 

If $P$ is a single Pauli operator, then $6(N-1)$ terms are flipped, so the energy is $12J \frac{N-1}{N}$. There are $3N$ such states.

A general Pauli string $P^1$ with $M$ non-identity operators corresponds to $M$ excitations. Each vector of the form $P^1 |\infty\rangle$ is an exact energy eigenstate of $H_1$,
\begin{equation}
    H_1 P^1 |\infty\rangle = E_1(P) P^1 |\infty\rangle.
\end{equation}
Furthermore, the energy eigenvalue is approximately additive, \begin{equation}
    E_1(P)\approx 12J M + \mathcal O(M/N),
\end{equation}
provided the excitations are dilute, $M \ll N$.

\subsubsection{General $k$}

In the general case, we have $k!$ ground states. The different pairings are not quite orthogonal at finite $N$, but they still span a $k!$-dimensional space of exact ground states. Again, this is true provided $k < D+1$.

The smallest gap arises from excitations on top of a single pairing. The spectrum is exactly known and matches the $k=1$ case above with the minimum gap $12J + \mathcal O(1/N)$. 

The other possible excitations are multiple excitations above a fixed pairing and domain walls. 
When the excitations are dilute, their energies approximately add. 
The domain wall state has energy $18J$ for even a single site in the domain, and the energy of the domain wall is extensive in the size of the domain. Hence, the whole partition function in the long time limit can be estimated by studying dilute towers of excitations on top of the $k!$ ground states. We present a detailed analysis of this situation for $k = 2$ in Appendix \ref{append:spingroundstates}.

\subsection{Path integral approach} 

Besides the effective Hamiltonian approach, the frame potential allows a path integral representation, and for our large $N$ model, a saddle point analysis. 
We present the saddle point analysis in the following. 
In this and the following sections, we use $T=2t$ to denote the evolution time for simplicity. 

Using spin coherent states, a time evolution operator can be cast into a path integral,
\bea
    && \tr[ U(T) ^{\otimes k} \otimes U(T)^{*\otimes k} ] \nn \\
    && = \int D\Omega\exp  \int dt \sum_r \Big[ \sum_{a=L,R;i} \langle \partial_t \Omega_{a,i}^r | \Omega_{a,i}^r \rangle \nn \\
    && - \sum_{i<j} \sum_{\alpha\beta} i J_{ij\alpha\beta} (\sigma_{L,i\alpha}^r \sigma_{L,j\beta}^r - \sigma_{R,i\alpha}^{r} \sigma_{R,j\beta}^{r}) \Big],
\eea
where $\langle \partial_t \Omega_{i}^r | \Omega_i^r \rangle$ is a short-hand notation for spin path integral, and $\Omega$ is a representation for the spin coherent state (such as Euler angle representation). The $L$ and $R$ subindices denote the fields in $U(T) = \mathcal T e^{i\int dt H(t)}$ and $U^\ast(T) = \mathcal T e^{i\int dt H^\ast(t)}$ respectively. $\mathcal T$ denotes time ordering.  
More precisely, $\sigma_{L,i\alpha} = \langle \Omega| {\bm \sigma}_{L,i,\alpha} | \Omega \rangle $ and $\sigma_{R,i\alpha} = \langle \Omega| \bm \sigma_{R,i,\alpha}^\ast | \Omega \rangle$, where $\bm \sigma$ is the Pauli operator. 
$r=1,...,k$ is the replica index. 

To get the frame potential, we average over the ensemble by integrating out the Brownian Gaussian variables,
\bea
    && F_{\text{b-spin}}^{(k)} (T) = \int D\Omega \exp \int dt \Big[ \sum_{r,a,i} \langle \partial_t \Omega_{a,i}^r | \Omega_{a,i}^r \rangle \nn \\
    && +  \frac{J}{4N} \sum_{r,s} \sum_{a\ne b} (\sum_{i\alpha} \sigma_{a,i\alpha}^r \sigma_{b,i\alpha}^s)^2\Big] - \frac{9}2 k N J T ,
\eea
where we use that $ \langle \Omega|\sum_{i} \bm \sigma_{a,i} \cdot \bm \sigma_{a,i} | \Omega \rangle = 3N$ for the spin coherent state to get the last term. 

To proceed, we introduce the Green's function $G^{rs}_{ab}(t) = \frac1N \sum_{i\alpha} \sigma_{a,i\alpha}^r(t) \sigma_{b,i\alpha}^s(t)$. 
After some manipulation (see Appendix~\ref{append:spin} for the derivation), the frame potential is given by
\bea
    && F_{\text{b-spin}}^{(k)} (T) = \int DG e^{-I}, \\
    && - \frac{I}{N} = \log \Tr e^{-\int dt H(t)} - \frac{J}4 \sum_{rs} \sum_{a\ne b} \int dt (G_{ab}^{rs})^2 - \frac92 k J T, \nn \\
\eea
where the first term in the large-$N$ action gives the free energy of $2k$ spin with the following Hamiltonian,
\bea
    H(t) &=& \frac{J}2 \sum_{rs} \sum_{a\ne b} G_{ab}^{rs}(t) \bm \sigma_{a}^r \cdot \bm \sigma_{b}^s,
\eea
where $\bm \sigma_{a}^r$, $a=L,R$, and $r,s=1,...,k$, denotes the Pauli operator for the $2k$ spins.

It is illuminating to first look at a single replica $k=1$. 
To look for a steady solution, we assume $G_{ab}$ is a constant (we omit the superscript for the replica index for the single replica case).
The Hamiltonian is 
\bea
    H = J G_{LR}\bm\sigma_{L} \cdot \bm\sigma_{R},
\eea
which only involves two spins $a= L, R$. Here we use the property $G_{ab} = G_{ba}$ that should be satisfied for a solution.
The eigenstates of this Hamiltonian include one singlet state and three triplet states. 
So the action becomes
\bea
    - \frac{I}N &=& \log [e^{ 3 G_{LR} J T } + 3 e^{-G_{LR} JT }] 
     -    \frac{JT}2 (G_{LR})^2  - \frac92 JT. \nn \\
\eea
In the long time limit, the saddle-point solution is $G_{LR} = 3$, consistent with the assumption of a time independent solution. 
Using this saddle point solution, the frame potential for $k=1$ is 
\bea
    F^{(1)}_\text{b-spin}(T) \approx \big( 1 + 3 e^{-12 J T} \big)^N \approx e^{3N e^{-12 J T}}. 
\eea
The decaying mode is precisely given by the elementary excitation discussed in Section~\ref{sec:spin-hamiltonian}.


Now consider general $k$.
We again assume the solution is time independent to look for steady solutions.
The action can be simplified to
\bea
    - \frac{I}{N} &=& \log \Tr e^{- T H} - \frac{JT}2 \sum_{rs} (G_{LR}^{rs})^2 - \frac92 k J T. \nn \\
    H(t) &=& J \sum_{rs} G_{LR}^{rs} \bm \sigma_{L}^r \cdot \bm \sigma_{R}^s.
\eea

At the long time limit, the first term projects to the ground state of $H$, which denote is denoted as $|\Psi\rangle$.
The equation of motion becomes
\bea \label{eq:spin-eom}
    G_{LR}^{rs} = - \langle \Psi | \bm \sigma_{L}^r \cdot \bm \sigma_{R}^s | \Psi \rangle. 
\eea
It is clear that for any state $| \Psi \rangle$, $- \langle \Psi | \bm \sigma_{L}^r \cdot \bm \sigma_{R}^s | \Psi \rangle \le 3$. 
The equality is saturated for a spin singlet state between $\bm \sigma_{L}^r$ and $\bm \sigma_{R}^s$, so we have $G_{LR}^{rs} \le 3$. 

Since the first term projects to the ground state, i.e., $\log \Tr e^{- T H} = - T \langle \Psi | H | \Psi \rangle $, using (\ref{eq:spin-eom}) the action becomes
\bea \label{eq:spin-onshell}
    - \frac{I}{N} & \rightarrow & \frac{JT}2 \sum_{rs} (G_{LR}^{rs})^2 - \frac92 k J T.
\eea
We expect (\ref{eq:spin-onshell}) to be time independent. 
As the second term is linear in the replica number, it indicates that the $L$ spins and the $R$ spins will form $k$ singlets. 
The choice of singlets state can be described by a permutation.
It is not hard to verify that at long time limit, the saddle point solution is
\bea
    G_{LR}^{rs} = 3\cdot [P(\pi)]^{rs},
\eea
where $P(\pi)$ is the permutation matrix corresponding to the permutation $\pi$. 

Indeed, this is consistent with the Hamiltonian analysis in Section~\ref{sec:spin-hamiltonian}. 
The singlet between the $L$ spin and $R$ spin is corresponding precisely to the EPR state.  

To get the finite time correction, we can plug the solution to the finite time action~(see Appendix~\ref{append:spin} for details), and by including the degeneracy from the $k!$ different permutation matrices $P(\pi)$, the frame potential is given by
\bea
    F_\text{b-spin}^{(k)} (T) \approx k! e^{-I} = k! e^{3 N k e^{-12 JT}}. 
\eea
Using this result, the time to get an approximate $k$-design reads
\bea
    t \ge \frac1{24J} \left( 3(\log2 + e^{-1}) kN + 2 \log \epsilon^{-1}_{\diamond} \right). 
\eea

\section{Brownian SYK Fermion Cluster}
\label{sec:brownianfermions}

The $k$-fold channel~(\ref{eq:k-channel}) averaged over Haar-distributed unitaries can be regarded as a projection. 
The invariant states for the projection are $|W_\pi \rangle  = \sum_{i_1,i_2,...,i_k} \bigotimes_{j=1}^k | i_j \rangle \otimes |\bar{i}_{\pi(j)} \rangle$, where $\pi$ is a permutation~\cite{roberts2017chaos}. 
The frame potential is to count the number of invariant states (see Appendix~\ref{append:fermiframepotential} for details), and since there are $k!$ different permutations, $F_\text{Haar}^{(k)}=k!$. 

For the unitary generated by the Hamiltonian of a fermionic model, it preserves a Fermi parity symmetry. Namely, $[U, (-1)^F]=0$. It means that there are more invariant states, given by the following ones, 
\bea
    |W_\pi^{\eta} \rangle = \sum_{i_1,i_2,...,i_k} \bigotimes_{j=1}^k | i_j \rangle \otimes (-1)^{\eta_j F}|\bar{i}_{\pi(j)} \rangle
\eea
where $\eta_j = 0,1 $. There are in total $2^k k!$ different invariant states. 
The frame potential will be $F_\text{f-Haar}^{(k)}  = 2^k k!$, where $F_\text{f-Haar}^{(k)}$ indicates the ensemble preserves Fermi parity symmetry. 
We expect the frame potential of the Brownian SYK model will be given by the same number.  


\subsection{Path integral approach}

The Brownian SYK model is defined as
\bea
	H(t) =  \sum_{i<j<k<l} J_{ijkl}(t) \psi_{i} \psi_{j} \psi_k \psi_{l} ,
\eea
where $\psi_{j}$, $j=1,...,N$, $\{ \psi_i, \psi_j \} = \delta_{ij}$ are Majorana fermions. $J_{ijkl}(t)$ is a Brownian Gaussian variable with mean zero and variance
\bea
	\avg{[J_{ijkl}(t)J_{i'j'k'l'}(0)]} = \delta_{i i'} \delta_{j j'} \delta_{k k'} \delta_{l l'} \frac{2^{3} 3! J}{N^{3}} \delta(t).
\eea

Similar to the Brownian spin model, we can formulate a path integral representation for the Brownian SYK model~(see Appendix \ref{append:syk} for detailed derivation). 
Introducing the bilocal variables $G_{ab}^{rs}(t,t') = \frac1N \sum_{i} \psi_{a,i}^r(t) \psi_{b,i}^s(t')$, and the associated self-energy $\Sigma_{ab}^{rs}$, where $a,b=L,R$ denote the field in $U$ and $U^\ast$ and $r,s=1,...,k$ are the replica indices, the frame potential is given by
\bea \label{eq:bSYK-action}
    && F_{\text{bSYK}}^{(k)}(T) = \int DG D\Sigma \exp N \Big[ \frac12 \log \det (\partial_t + \Sigma) \nn \\
    && + \int dt dt' \left( \frac12 \Sigma_{ab}^{rs} G_{ab}^{rs} +  \frac{J \delta(t-t')}{16} c_{ab} (2  G_{ab}^{rs})^4  \right),
\eea
where $c_{LL} = c_{RR} = -1$, $c_{LR} = c_{RL} = 1$, and the sum over indices $a,b,r,s$ is implicit. 

Inspired by the Brownian spin model, we expect the steady state is formed by paring one forward $U$ and one backward evolution $U^\ast$.
Without loss of generality, we assume the $r$-th and $s$-th replica have nontrivial correlations. 
Under this assumption, the equation of motion reduces to a two by two matrix equation, and can be solved exactly (see Appendix~\ref{append:syk} for details). The solution is given by 
\bea \label{eq:rs-solution}
     \hat G(t_1,t_2) &=&   \Big[  \frac{\theta(t_{12})}2 f_+(t_{12}) - \frac{\theta(t_{12})}2 f_+(-t_{12}) \Big] \bm 1 \nn \\
    && \pm \Big[\frac{\theta(t_{12})}2 f_-(t_{12}) - \frac{\theta(t_{12})}2 f_-(-t_{12}) \Big] \bm \sigma_2,  \\
     \hat \Sigma(t_1, t_2) &=& \pm \delta(t_{12}) J \bm \sigma_2,
\eea
where $ \hat G = \left( \ba{cccc} G_{LL}^{rr} & G_{LR}^{rs} \\ G_{RL}^{sr} & G_{RR}^{ss} \ea \right)$ and $t_{12} = t_1 - t_2$, and $ f_{\pm}(t) = \frac{ e^{- J t}}{e^{-J T} + 1} \pm \frac{ e^{J t}}{e^{J T} + 1}$.
$\bm 1$ is a two-by-two identity matrix and the Pauli matrix $\bm \sigma_2$ is a Pauli matrix. 
The plus and minus sign means that there two solutions for a fixed $rs$.

Because this solution is decoupled from the rest, we can calculate the onshell action for a fixed $rs$ (see Appendix~\ref{append:syk} for details), 
\bea \label{eq:slowestspin}
    - \frac{I_{\text{wh}}(T)}{N} = \log \left( 2\cosh\frac{J T}2 \right) - \frac{JT}2,
\eea
where the subscript $\text{wh}$ indicates that it is like a wormhole solution connecting the $r$-th and $s$-th replicas. 
On the other hand, there is an obvious trivial solution given by $\hat G = [\sgn(t_{12})/2] \cdot \bm 1 $, $\hat \Sigma = 0$, with onshell action  
\bea
- \frac{I_{0}(T)}{N} = \log2 - \frac{JT}8, 
\eea
where 
the subscript $0$ indicates that it comes from a trivial solution. 

Thus, when there is a nonvanishing correlation between them, the onshell action tends to zero at long time limit. 
On the other hand the trivial solution tends to infinity at the long time limit, indicating that they do not contribute to the frame potential: it is exponentially suppressed. 
While at time zero, both solutions leads to the dimension of Hilbert space, $2^N$. 
This number is because each replica contains $N$ Majorana fermoins.

The above solutions are restricted to a two by two matrix, i.e., the $r$-th and $s$-th replicas, and the remaining step is to count how many different solutions can exist in $k$ replicas. 
The result is given by
\bea \label{eq:k-saddle}
	F_{\text{bSYK}}^{(k)}(T) 
	&=&  \sum_{m=0}^k 2^m m! \left(\ba{cccc} k \\m \ea\right)^2 e^{-(k-m)I_0(T)} e^{- m I_\text{wh}(T)}, \nn \\
\eea
where $m$ indicates the number of nontrivial paired solutions and $k-m$ indicates the number of trivial solutions in the $k$ possibilities. 
The prefactor is the degeneracy. 

At time zero, the frame potential should give the dimension of Hilbert space of $2k$ Brownian SYK systems. 
It seems that our result is greater than the dimension of Hilbert space because the contribution from nontrivial solutions are included. 
But as we will see in the following, those nontrivial solutions should not be included at time zero. 

At the long time limit, the trivial solution will lead to an exponential suppression with exponent proportional to $N$, i.e.,
\bea
    e^{-(k-m)I_0(T)} = e^{(k-m) N (\log 2 - \frac{JT}8)}, 
\eea
for $m<k$. 
Thus, the dominant solution is given by $m=k$ with maximal pairs in the $k$ different replicas, 
\bea \label{eq:syk-potential}
    F_{\text{bSYK}}^{(k)}(T) 
    &\approx& 2^k k! (1+ k N e^{- J T}).
\eea
Regarding the prefactor, while $2^k$ is given by the sign choice in~(\ref{eq:rs-solution}), $k!$ corresponds nicely to the permutation. 


As we will see in the next section, the slowest decaying mode is actually given by an elementary excitation, i.e., a single Majorana fermion. 
But one may wonder if collective modes can have a smaller gap. 
To see it is not the case, we consider fluctuations around the saddle point solution. Schematically, we consider the fluctuation,
\bea
G = \bar G + \frac1{\sqrt N} g, \quad \Sigma = \bar \Sigma + \frac1{\sqrt N} \sigma.
\eea
Here, we use $\bar G$, $\bar \Sigma$ to denote the saddle point solution, and $g$, $\sigma$ to denote the fluctuation, and the prefactor is a proper scaling for a large $N$ theory.

It turns out that we can focus on the fluctuations around (\ref{eq:rs-solution}) for each fixed $rs$ (we omit the index $rs$ in the following for simplicity). 
We expand the action to the quadratic order of both $\sigma$ and $g$, and then integrate out $\sigma$ to arrive at (see Appendix~\ref{append:correction} for details)
\bea \label{eq:boson}
    && \delta I = \nn \\
    && \frac1{2T} \sum_\omega \hat g(-\omega) \left( \ba{cccc} 2J \coth \frac{JT}2 & -\omega \coth \frac{JT}2 \\
    \omega \coth \frac{JT}2 & 2J \coth \frac{JT}2 + 12 J \ea \right) \hat g(\omega), \nn \\
\eea
where $\hat g = (g_{LL}, g_{LR})^T$, while the other two components are related by $g_{RR} = g_{LL}$ and $g_{RL} = g_{LR}$. 
Now it is a free boson, we can get its free energy $\delta F = \frac1T \log (1 - e^{- E T})$,
where $E = 2J \sqrt{6 \tanh \frac{JT}2 - 1}$ is the gap of the boson. 
There is a critical $T^*$, i.e., $ \tanh \frac{JT^*}2 = \frac16 $, after which the boson becomes stable. 
When $T<T^*$, we should not include the wormhole contributions, so the frame potential (\ref{eq:k-saddle}) at time zero is correctly given by the dimension of Hilbert space.  

For a fixed permutation solution, we have $k$ different pairs of $rs$, which leads to $k$ bosons. 
Because we are interested in the long time behavior, we expect that a renormalization flow of the mass $E \approx 2\sqrt5 J$ due to high order corrections gives an appropriate estimate of the collective decaying mode. 
The large $N$ structure indicates that correction is of order $\mathcal{O}(\frac1N)$. 
So we expect the free energy and thus the frame potential is given by
\bea
    &&F_{\text{bSYK}}^{(k)}(T) = 2^k k! e^{-FT}, \\
    && - F T \approx k N \log (1+ e^{- J T}) - k \log(1 - e^{- \big(E + \mathcal{O}( \frac{1}N ) \big) T}), \nn \\
\eea
where the first term is from elementary excitation, and the second term is from collective modes.

Using this result, the time to reach an approximate $k$-design at large $N$ limit is given by \bea
    t \ge \frac1{2J} \left( (\frac32 \log2 + e^{-1}) N k + 2 \log \epsilon_\diamond^{-1} \right).
\eea
Note that here $k \le 2^{N/2}$ since the Hilbert space dimension is $D = 2^{N/2}$ for $N$ Majorana fermoins.

\subsection{Effective Hamiltonian}
Here we provide an effective Hamiltonian analysis, and identify the elementary excitation given rise to the slowest decaying mode. 
For the Brownian SYK model, the Hamiltonian is given by~(see Appendix \ref{append:syk-hamiltonian} for details),
\bea
    && H = \frac{4 \cdot 3! J}{ N^{3}} \times \nn \\
    && \sum_{ijkl} \big[\sum_r ( \psi_{L,i}^r \psi_{L,j}^r \psi_{L,k}^r \psi_{L,l}^r - \psi_{R,i}^r \psi_{R,j}^r \psi_{R,k}^r \psi_{R,l}^r) \Big],
\eea
where $r=1,...,k$ is the replica index. 
Similar to the Brownian spin model, the ground state of this Hamiltonian is given by the tensor product of EPR state, each of which is from two possible EPR states (distinguished by the plus and minus signs in the following) between two contours $\alpha$ and $\alpha'$
\bea
   && |\text{GS} \rangle = \bigotimes_{r,s} |\infty \rangle_{r,s}, \\ 
    && (\psi^{r}_{L,j} \pm i \psi^{s}_{R,j} )| \infty \rangle_{r,s} = 0, \quad \forall j=1,...,N.
\eea
where in the tensor product $r$ and $s$ can only appear once. 
The number of such ground states is given by 
\bea
    N(\text{GS}) = 2^k k!.
\eea

The elementary excitation (eigenstate) is given by $\psi_{L,i}^r|\text{GS} \rangle $ for any flavor $i$ and replica $r$. The energy of this excitation is
\bea
    H \psi_{L,i}^r|\text{GS} \rangle = J \psi_{L,i}^r|\text{GS} \rangle.
\eea
The number of the elementary excitation is given by $kN$ for each degenerate ground state, so we expect the frame potential is given by
\bea
    \label{eq:framepotbsyk}
    F_{\text{bSYK}}^{(k)}(T) = 2^k k! (1 + k N e^{-JT}),
\eea
which exactly reproduces the saddle point calculation in~(\ref{eq:syk-potential}). One should also notice that the swap operator in the fermionic case is not an independent mode but simply given by the elementary excitation.

\section{Time-independent Hamiltonians}
\label{sec:timeindepham}

Whereas the time-dependent Brownian models considered above are especially convenient for theoretical analysis, similar tools can also be used to characterize the growth of complexity in systems governed by time-independent Hamiltonians such as those describing strongly-interacting chaotic many-body quantum systems. To make analytical progress in this scenario, we perturb the many-body Hamiltonian $H$ by Brownian time-dependent sources, which allows us to map the problem onto an effective statistical model after averaging over the Brownian noise. Specifically, we consider perturbing the original Hamiltonian $H$ by time-dependent sources $O_{\alpha}$:
\begin{equation}
    H \rightarrow H + \sum_\alpha \xi_\alpha(t) O_\alpha.
\end{equation}
where the sources $O_{\alpha}$ are Hermitian and square to the identity $O_{\alpha}^2 = \id$, and the couplings $\xi_\alpha(t)$ are white-noise random variables with mean zero and variance $\avg{(\xi_{\alpha}^2)} = g/dt$ for an infinitesimal time step $dt$. 
Driving the system in this way typically adds energy to the system, so we expect the system to heat up to infinite temperature at long times. It is interesting to consider whether variations of this model might allow us to ask similar questions at finite temperature, but we leave these possibilities open for future work.


To characterize the growth of complexity in this system, we again compute the $k$-th frame potential, which involves two copies of the operator
\begin{equation}
    M_U = \underbrace{U \otimes \cdots \otimes U }_{\text{$k$ times}} \otimes \underbrace{U^* \otimes \cdots U^*}_{\text{$k$ times}}.
\end{equation}
We have transposed the inverse time-evolution operators $U^\dagger$s to guarantee that all copies of each random $\xi_\alpha(t)$ occur at the same time slice. The FP is then
\begin{equation}
    F^{(k)}_H = \avg_{U,V}[ \tr( M_V^\dagger M_U)] = \tr( [\avg_U M_U]^\dagger [\avg_U M_U] ).
\end{equation}
Because the expectation values factorize, we simply need to compute $\avg_U M_U$ and study the decay rate of an initial state summed over all initial states.


Because the Brownian coefficients $\xi_{\alpha}(t)$ are uncorrelated at different times, we can compute the disorder average separately for each timestep $dt$, similar to above. Performing this disorder average
for the full $2k$ copies of $U,U^*$ after evolving for time $T$ yields an effective time-evolution operator $\exp\left[ - i H_{\text{eff}} T \right]$
where
\begin{equation}
    \label{eq:effhamrmt}
    H_{\text{eff}} = \sum_r \{ H^r  - H^{\bar{r}}\} - \frac{i g}{2} \sum_\alpha \left( \sum_{r=1}^k \{ O_\alpha^{r} - (O_\alpha^{\bar{r}})^*\}\right)^2.
\end{equation}
Notice that this effective Hamiltonian is non-Hermitian due to the intrinsic decay from the $O^2$ terms. This is distinct from the completely Brownian model considered above, which produced entirely imaginary terms in the effective Hamiltonian, allowing us to reinterpret the whole problem as an equilibrium calculation in imaginary time. In the present case, the imaginary terms are perturbations to a fixed Hermitian Hamiltonian $H$, so instead we think of the non-Hermitian $O^2$ term as driving the decay of eigenstates of the Hamiltonian $H$.


At long times $T \rightarrow \infty$ we expect all states to completely decay, except for a privileged set of at least $k!$ `dark states' that never decay. These states correspond to all possible pairings of forward and backward contours into $k$ infinite temperature thermofield double states (see Appendix \ref{append:nonhermitianperttheory}). Integrable systems with further symmetries may have additional dark states. To preempt this possibility, we here consider only random matrix Hamiltonians with strong level repulsion.

\begin{figure}
    \centering
    \includegraphics[width=.8\columnwidth]{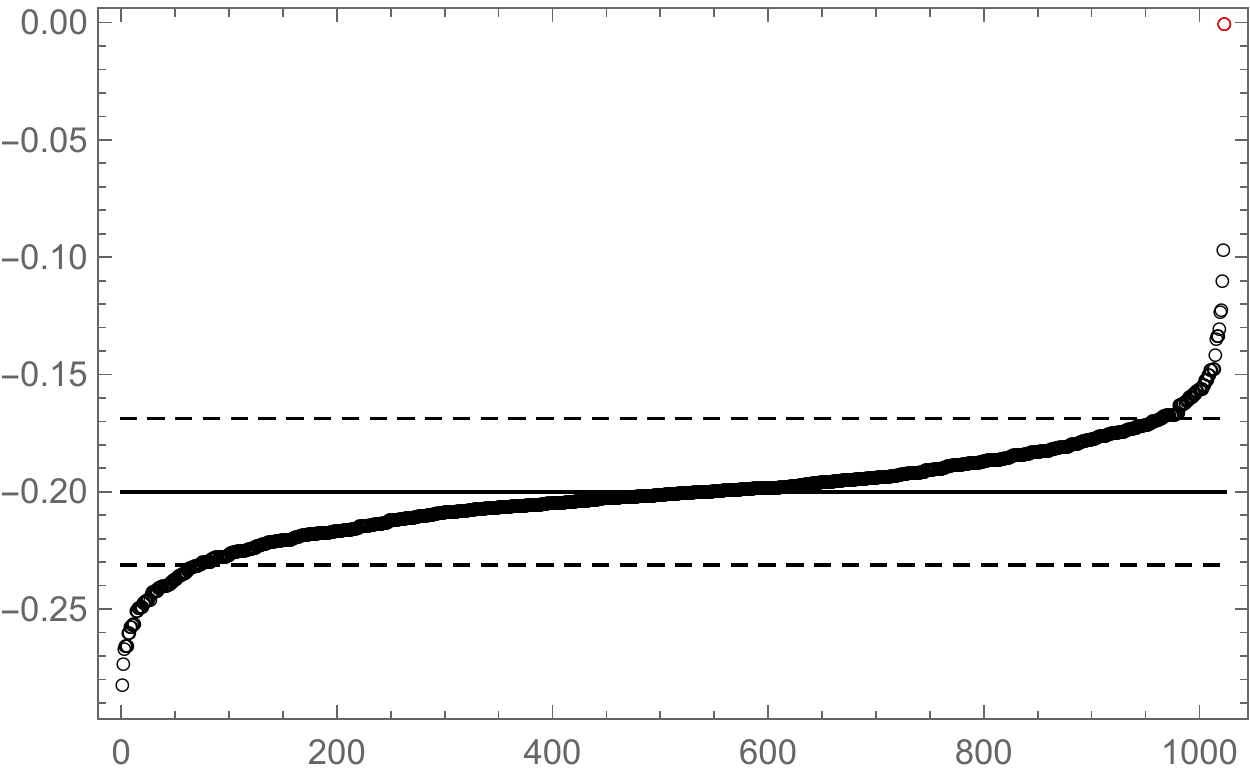}
    \caption{Imaginary part of eigenvalues of $H_{\text{eff}}$ for $D=32$ and $k=1$ with $H$ a GOE random matrix, a single $O$ given by a diagonal matrix with equal numbers of $\pm1$s, and $g=0.2$. The standard deviation of terms in $H$ is taken to be $
    \log(D)/\sqrt{D}$ to mimic an extensive qubit Hamiltonian with spectrum in $[-\log(D),\log(D)]$. The dashed horizontal lines indicate $1/D$ variance around $-g$, and the rightmost red circle indicates the dark state.}
    \label{fig:goe_spectrum}
\end{figure}

Assuming that all other states decay, it follows that at infinite time the FP will be $k!$, the Haar value. Hence, the perturbed ensemble will eventually form a $k$-design provided these thermofield doubles are the only dark states. Below, we demonstrate this in an explicit example where $H$ is a random matrix chosen from the GOE ensemble.

\subsection{Random Matrix Hamiltonians}

For simplicity, consider choosing the Hamiltonian $H$ from a standard random matrix ensemble, say the GOE ensemble. For simplicity, we consider a single perturbing operator $O_\alpha = O$ that is diagonal in the computational basis with an equal number of $\pm1$ randomly placed along the diagonal. We can numerically study the spectrum of the resulting non-Hermitian $H_{\text{eff}}$ for $k=1$. We find that there is indeed just $1!=1$ dark state $\ket{d}$ and that all other states decay.  An example of the spectrum can be seen in Fig.~\ref{fig:goe_spectrum}; note the single dark state at the top of the graph.

We can readily derive a perturbation theory argument for this same conclusion. Consider first the case $k=1$ and regard the non-Hermitian $O^2$ term as a perturbation on the bare Hamiltonian $H$. The bare eigenstates are $|n,m\rangle$ with energies $E_n - E_m$. Non-Hermitian perturbation theory (see Appendix \ref{append:nonhermitianperttheory}) shows that the first correction to the energy of the $|n,m\rangle$ state is
\begin{align}
    \label{eq:decayratenondegen}
    \delta E &= - i \frac{g}{2} \langle n,m| \left(O^{1}-O^{\overline{1}}\right)^2 |n,m\rangle \nonumber \\
    &= - i g (1 - \langle n,m| O^{1} O^{\overline{1}} |n,m\rangle) \nonumber \\
    &\approx - i g (1 + \mathcal{O}(1/D)),
\end{align}
since the matrix elements of the fixed operator $O$ in the random basis obtained from $H$ will be of order $1/\sqrt{D}$. Hence, first order perturbation theory immediately predicts a constant decay rate $g$ for all states, up to corrections of order $1/D$, in agreement with the spectrum shown in Fig. \ref{fig:goe_spectrum}.

\begin{figure}
    \centering
    \includegraphics[width=.8\columnwidth]{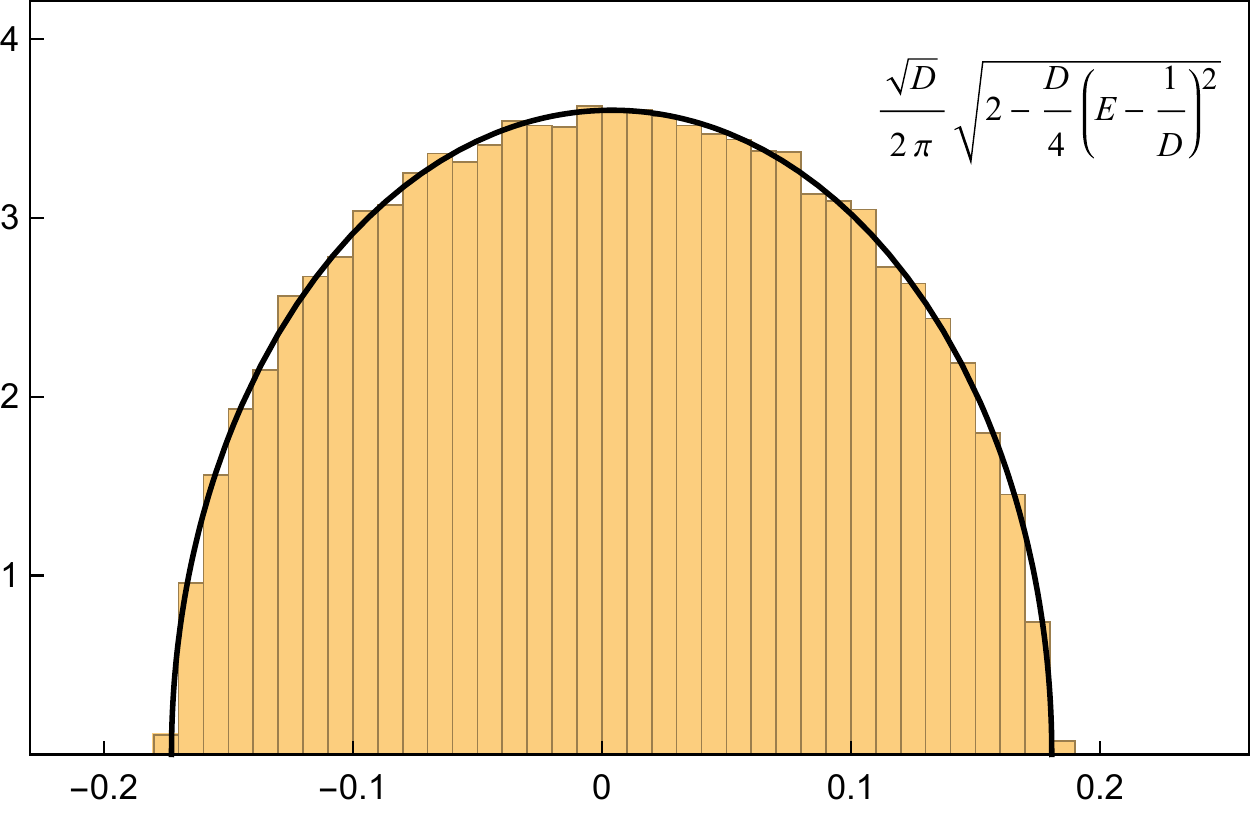}
    \caption{The distribution of eigenvalues of $M$. The figure shows exact diagonalization for $D=256$ from $50$ samples. The total density is normalized to one.}
    \label{fig:semicircle}
\end{figure}

The special case $n = m$, however, requires more care because the bare energy is degenerate $E_n - E_n = 0$. The matrix elements of the perturbation in the degenerate space are 
\begin{align}
    \label{eq:decayratedegen}
    - &i \frac{g}{2} \langle n,n| \left(O^{1}-O^{\overline{1}} \right)^2 |m,m\rangle \nonumber \\
    = - &ig (\delta_{n,m} - \langle n |O|m\rangle \langle m |O|n\rangle).
\end{align}
where we have used $O^{\bar 1} = O^T$. 
The dark state $\ket{d}$ is in this sector, and is given by a symmetric superposition over all basis states $\ket{n}_1$ and their time reversed copies $\ket{n}_{\overline{1}}$:
\begin{equation}
    \label{eq:darkstate}
    \ket{d} = \frac1{\sqrt D} \sum_n \ket{n,n}_{1 \overline{1}}.
\end{equation}

Accounting for this dark state, we now look at other eigenstate of~(\ref{eq:decayratedegen}).
The matrix elements take the form (since the identity matrix is simple, we consider the nontrivial part in (\ref{eq:decayratedegen}))
\begin{equation} \label{eq:perturb-matrix}
    M_{nm} = |\langle n | O |m \rangle|^2, 
\end{equation}
where we have used the fact that $O$ is a Hermitian operator. 
It can be shown that the matrix element satisfies the Porter-Thomas distribution with mean $1/D$ and variance $2/D^2$~(see Appendix~\ref{append:perturb-matrix}).
Because the matrix $M$ is symmetric, and upper-triangular entries satisfy identical independent distribution, its eigenvalues satisfy the Wigner's semicircle law~\cite{brody1981random},
\bea
    \rho(E) = \frac{\sqrt{D}}{2\pi} \sqrt{2- \frac{D}4 
    \left(E- \frac1D \right)^2},
\eea
showing that the spectrum of the random matrix $M$ will sit between $[1/D- \sqrt{8/D}, 1/D+ \sqrt{8/D}]$. 
It is consistent with exact diagonalization of $M$ in Fig.~\ref{fig:semicircle}. 

\begin{figure}
    \centering
    \includegraphics[width=.8\columnwidth]{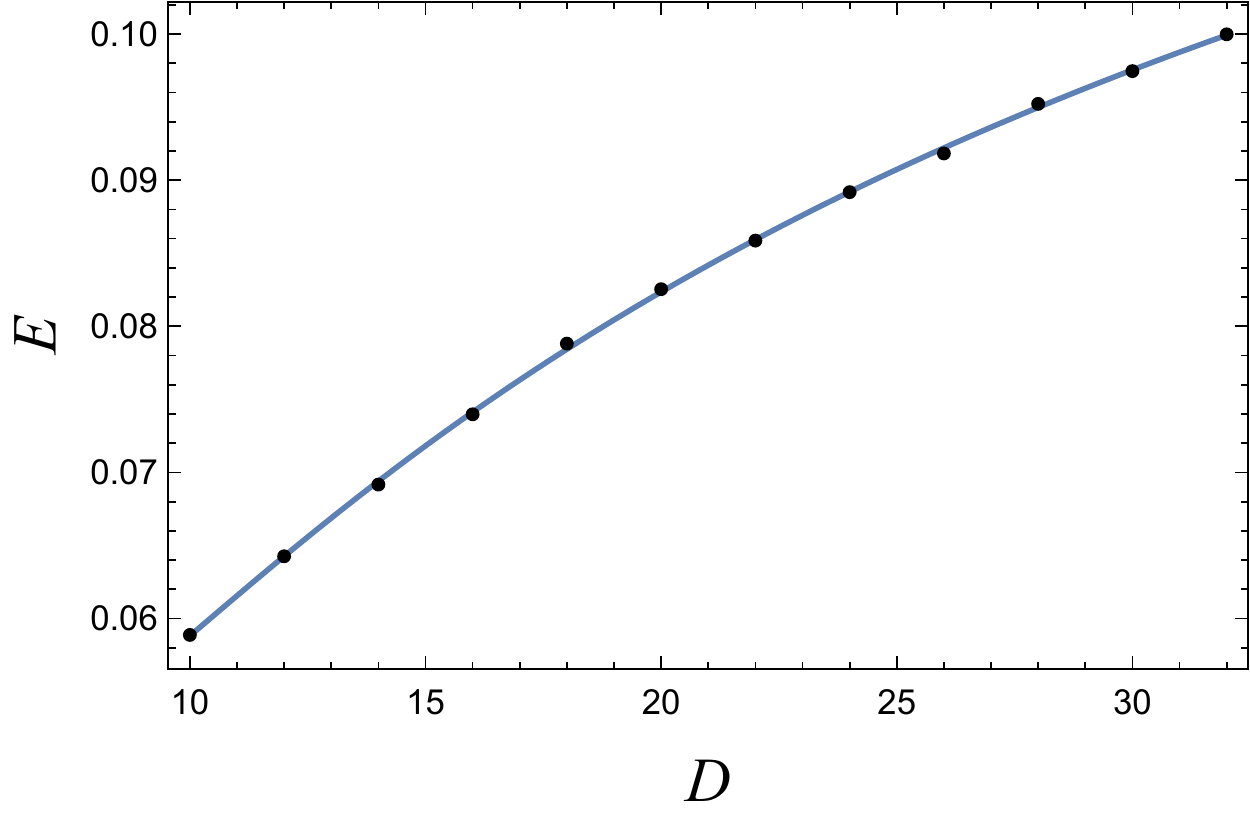}
    \caption{Average decay gap of $H_{\text{eff}}$ for $k=1$ as a function of Hilbert space dimension $D$. $H$ is a GOE random matrix, $O$ is a diagonal matrix with equal numbers of $\pm1$, and $g=0.2$. The standard deviation of terms in $H$ is taken to be $
    \log(D)/\sqrt{D}$. For $D=10,12...,24$ we average over 2000 samples, and for $D=26,...,32$ we average over 1000 samples. The blue curve is a fit to function $a + b/\sqrt{D} + c/D$, where the fit value of $a$ is $0.2006$ which is very close to $g=0.2$.}
    \label{fig:goe_decay_gap}
\end{figure}

This analysis then predicts a single dark state, and the rest states with decay rate $-i g $ with corrections of order $1/\sqrt{D}$. 
A fit of the minimum averaged decay rate (apart from the dark state) for even dimensions from $D=10$ to $D=32$ is shown in Fig.~\ref{fig:goe_decay_gap} 

In the case of general $k$, we can make similar perturbative arguments starting from a given pairing (see Appendix \ref{append:nonhermitianperttheory}). Hence, we expect $k!$ dark states and a minimal decay rate of order $g(1- \mathcal O(1/\sqrt D))$. The FP can then be bounded by
\begin{equation}
    \label{eq:framepotham}
    F^{(k)}_H \leq k! + (D^{2k}-k!)e^{-2g t}.
\end{equation}
So taking $D=2^N$ we get an approximate $k$-design after a time
\begin{equation}
     t > \frac1{2g} \left( 4 k N \log 2 + 2 \log \epsilon_\diamond^{-1} \right)
\end{equation}
growing linearly in the system size $N$. While the above analysis relies on the fact that $H$ is chosen uniformly from the GOE ensemble, we expect similar conclusions to hold when $H$ is any strongly-mixing Hamiltonian satisfying the eigenstate thermalization hypothesis. We leave a detailed study of this possibility to future work.

\section{Discussion and Outlook}
\label{sec:discussion}

In this paper we studied the quantum complexity growth problem in a variety of quantum chaotic models. 
While quantum complexity has multiple meanings and definitions, we focused here on $k$-designs, which measures how close an ensemble of unitaries is to the uniform distribution equipped with the Haar measure. 
We approach the problem by investigating the frame potential that quantifies the distance to a $k$-design. 
Compared to other measures of complexity, the frame potential provides a significant advantage that allows one to map the problem to a path integral representation which brings to bear many well-developed tools in physics including spectral and saddle-point analysis. 
These tools enable us to prove the linear growth of circuit complexity in our models.

Our models give representatives from different aspects of quantum complexity: the Brownian spin model directly links to random quantum circuits, the Brownian SYK model gives insight into holographic complexity, and the random matrix model is a useful representation of chaotic Hamiltonians. 
More importantly, our approach actually provides connections between these complementary approaches. 
In all three descriptions, the $k$-th frame potential involves $2k$ replicas. 
On the one hand, the saddle point analysis in the Brownian spin/SYK model clearly reveals that the saddle point solutions are classified by permutations, which is in accordance with the effective statistical model arisen in random circuits~\cite{zhou2019emergent,hunter2019unitary}. 
Moreover, these solutions describe an emergent correlation between different replicas, resembling the replica wormhole geometry from the gravitational path integral \cite{penington2022replica, almheiri2020replica}. 
On the other hand, the effective Hamiltonian approach in the random matrix Hamiltonian allows us to bring the full machinary of random matrix theory. 
While it is true that time independent Hamiltonians are not able to generate a random uniform ensemble, adding a simple time-dependent perturbation are,  provided the Hamiltonian exhibits eigenstate theormalization hypothesis. 
All these concepts are closely interrelated in quantum chaos. 

Our studies provide a firm base for wide future directions. 
Looking forward, an outstanding question is the holographic dual of the quantum complexity. 
Yet the conjectured complexity-volume and complexity-action dualities \cite{brown2016complexity,brown2016holographic,susskind2016computational,stanford2014complexity} still remain vague, our work creates a new train of thought.
As the replica wormhole underscores quantum thermalization in gravity, such as the von Neumann entropy and the spectral form factor, it seems natural to have a similar story in a holographic dual of the frame potential. 
In a broader context, how thermalization and complexity are related? 
It would be interesting to look into the complexity growth in generic quantum chaotic Hamiltonians, as our work suggests that upon a simple Brownian perturbation, those Hamiltonians are able to produce a linear growth.



\section*{Acknowledgement}

We thank Nick Hunter-Jones and Jonas Haferkamp for useful discussions. SKJ is supported by the Simons Foundation via the It From Qubit Collaboration.
GB is supported by the DOE GeoFlow
program (DE-SC0019380).
The work of BGS is supported in part by the AFOSR under grant number FA9550-19-1-0360.

\bibliography{refs}

\appendix


\section{Frame potential with Fermi parity symmetry}
\label{append:fermiframepotential}

Averaging over Haar measure of the unitary group can be regarded as a projection to invariant states.
For the $k$-fold channel, the invariant states are $|W_\pi \rangle =\sum_{i_1,i_2,...,i_k} | i_1\rangle \otimes |\bar{i}_{\pi(1)} \rangle \otimes ... \otimes  | i_{k}\rangle \otimes |\bar{i}_{\pi(k)} \rangle = \sum_{i_1,i_2,...,i_k} \bigotimes_{j=1}^k | i_j \rangle \otimes |\bar{i}_{\pi(j)} \rangle$, where $\pi$ is a permutation~\cite{roberts2017chaos}. Thus, there are $k!$ different invariant states and the average over Haar ensemble of the unitary group in the $k$-fold channel leads to
\bea
    \int dU U^{\otimes k } \otimes U^{\ast{\otimes}k} = \sum_{\sigma,\pi} [Q^{-1}]_{\sigma,\pi} | W_\sigma \rangle \langle W_\pi |, 
\eea
where $\int dU$ indicates the integral over Haar measure, and $Q_{\sigma,\pi} = \langle W_\sigma| W_\pi \rangle $. 
In this language, the frame potential is to count the number of invariant states. 
Namely,
\bea
    F_\text{Haar}^{(k)} &=& \int dU dV |\Tr(U V^\dag )|^{2k} \nn \\
    &=&  \int dU dV \Tr[U^{\otimes k} V^{\dag \otimes k}] \Tr[U^{\ast \otimes k} V^{T \otimes k}] \nn\\
    &=& \sum_{\pi_1,\pi_2,\pi_3,\pi_4} (Q^{-1})_{\pi_1,\pi_2} Q_{\pi_2,\pi_3} (Q^{-1})_{\pi_3,\pi_4} Q_{\pi_4, \pi_1} \nn \\
    &=& \sum_{\pi} 1 = k!,
\eea
where in the last equality, the summation over elements in permutation group gives $k!$.

For the unitary generated by the SYK models, it preserves the Fermi parity symmetry. 
Namely, $[U, (-1)^F]=0$. 
It means that there are more invariant states, as given by the following ones, 
\bea
    |W_\pi^{\eta} \rangle = \sum_{i_1,i_2,...,i_k} \bigotimes_{j=1}^k | i_j \rangle \otimes (-1)^{\eta_j F}|\bar{i}_{\pi(j)} \rangle
\eea
where $\eta_j = 0,1 $. There are in total $2^k k!$ different invariant states. 
Thus, the frame potential will be given by similarly summing over all the invariant states,
\bea
    F_\text{Haar}^{(k)} &=& \int' dU \int' dV |\Tr(U V^\dag )|^{2k} = 2^k k!,
\eea
where $\int ' dU $ denotes the Haar measure in the subspace that preserves Fermi parity symmetries.

\section{Ground States of Brownian Spin Effective Hamiltonian}
\label{append:spingroundstates}

In Section \ref{sec:spin-hamiltonian} of the main text we presented a detailed analysis of the ground-state manifold and gap to the excited state in the Brownian spin model for $k = 1$. Here we present a similar analysis for $k = 2$.


There are now $2!$ distinct ground states, and the permutation symmetry is spontaneously broken in each pairing of contours into thermofield doubles. 

We may consider excitations around any particular pairing. If only one of the pairings is excited, then the spectrum is identical to the $H_1$ case. 
The minimal energy cost of such an excitation, corresponding to a single Pauli operator, is $12J$. There are $6 N$ such states.

Now what if both pairs are excited? Consider an excitation of the form
\begin{equation}
    P^1 Q^2 |\infty \rangle_{1\bar{1}}|\infty \rangle_{2\bar{2}}
\end{equation} 
and a term
\begin{equation}
    \frac{J}{2N}\sum_{\alpha\beta}\left( \bm\sigma^{1}_{i\alpha} \bm\sigma^1_{j\beta} + \bm\sigma^{2}_{i\alpha} \bm\sigma^2_{j\beta} - (\bm\sigma^{\bar{1}}_{i\alpha} \bm\sigma^{\bar{1}}_{j\beta})^* - (\bm\sigma^{\bar{2}}_{i\alpha} \bm\sigma^{\bar{2}}_{j\beta})^*\right)^2.
\end{equation}
There are four possibilities, $P^1$ and $Q^2$ can commute or anticommute with the $O^1$ and $O^2$ operators in the term. If at most one of them anti-commute, then we're back in the situation with $H_1$. The term then contributes to the energy exactly like it would in the $H_1$ case.

If both anti-commute, then after moving $P^1$ and $Q^1$ through the term, the term becomes
\bea
      && \frac{J}{2N}\left( -\bm\sigma^{1}_{i\alpha} \bm\sigma^1_{j\beta} - \bm\sigma^{2}_{i\alpha} \bm\sigma^2_{j\beta} - (\bm\sigma^{\bar{1}}_{i\alpha} \bm\sigma^{\bar{1}}_{j\beta})^* - (\bm\sigma^{\bar{2}}_{i\alpha} \bm\sigma^{\bar{2}}_{j\beta})^*\right)^2, \nn \\
      && = \frac{J}{2N}\left( 2 \bm \sigma^{1}_{i\alpha} \bm\sigma^1_{j\beta} + 2 \bm\sigma^{2}_{i\alpha} \bm\sigma^2_{j\beta}\right)^2,
\eea
where we used operator pushing from $\bar{1}$ to $1$ and from $\bar{2}$ to $2$. 

The term no longer has $|\infty \rangle_{1\bar{1}}|\infty \rangle_{2\bar{2}}$ as an eigenstate, but we can compute its average in that state (since we already moved the Pauli strings through), giving
\begin{equation}
    2 \times 2^2 \times \frac{J}{2N}.
\end{equation}
This is a small correction on top of the $H_1$ energies of $P^1$ and $Q^2$ and does not close the gap.

Suppose $P$ and $Q$ are single Pauli operators. Then only an $\mathcal O(1)$ number of terms out of the $\mathcal O(N^2)$ total number of terms will anti-commute with both $P$ and $Q$. Since such terms can only change the energy by an $\mathcal O(1/N)$ amount, we conclude that the energy of two excitations is approximately additive.

There is one exception to this rule, which occurs when the excitations are on the same site in replica $1$ and $2$. In that case, every term that anti-commutes with $P$ also anticommutes with $Q$. In this special case, there are a $\mathcal O(N)$ terms that anti-commute with both (and no terms that anti-commute with only one), but the average energy of this state is still the same as for two decoupled excitations. Moreover, while the state is not an exact eigenstate, by computing the variance of the energy we learn that the energy is peaked around the average.

The general result is that for any dilute set of excitations, the leading contribution to the energy is the sum of the $H_1$ energies for those excitations taken in isolation. For an extensive (but dilute) number of excitations in each pairing, there will be an extensive (but small) correction to the total energy which vanishes as the density of excitations becomes more dilute.

A new type of excitation that is possible when $k>1$ is a ``domain wall'' between the two different ways of pairing the system to produce a zero-energy state. Suppose $N_1$ spins are paired as $1\bar{1},2\bar{2}$ and $N_2$ spins are paired as $1\bar{2},2\bar{1}$.

We can divide the $J$ terms into three sets, those that couple within $N_1$, those that couple within $N_2$, and those that couple between $N_1$ and $N_2$. The within terms are still exactly satisfied, so we need only consider the terms that couple $N_1$ to $N_2$. For each $i$ from the $N_1$ and each $j$ from the $N_2$, we have a term
\begin{equation}
    \frac{J}{2N}\sum_{\alpha\beta}\left( \bm\sigma^{1}_{i\alpha} \bm\sigma^1_{j\beta} +\bm\sigma^{2}_{i\alpha} \bm\sigma^2_{j\beta} - (\bm\sigma^{\bar{1}}_{i\alpha} \bm\sigma^{\bar{1}}_{j\beta})^* - (\bm\sigma^{\bar{2}}_{i\alpha} \bm\sigma^{\bar{2}}_{j\beta})^*\right)^2.
\end{equation}
When acting on the domain wall state, this term is equivalent to
\bea
    && \frac{J}{2N}\sum_{\alpha\beta}\left( \bm\sigma^{1}_{i\alpha} \bm\sigma^1_{j\beta} +\bm\sigma^{2}_{i\alpha} \bm\sigma^2_{j\beta} - \bm\sigma^{1}_{i\alpha} \bm\sigma^{2}_{j\beta} - \bm\sigma^{2}_{i\alpha} \bm\sigma^{1}_{j\beta} \right)^2 \nn \\
    && = \frac{2J}{N} \sum_{\alpha \beta} (1-\bm \sigma^1_{i\alpha}\bm\sigma^{2}_{i\alpha})(1-\bm\sigma^1_{j\beta}\bm\sigma^{2}_{j\beta}).
\eea
Now, the domain wall state is not an eigenstate of this term, but we can compute the expected value, $\frac{2J}{N}$.
There are $N_1 N_2$ terms (plus the sum over $\alpha\beta$), so the expected value of the energy is
\begin{equation}
    \frac{2 J N_1 N_2 3^2}{N} = 18 J \frac{N_1N_2}{N}.
\end{equation}

If we have a domain wall separating macroscopic domains, then the energy is extensive in $N$.
The minimal energy of such a domain wall is obtained for $N_1=1$ and $N_2=N-1$, in which case the energy is $18J$ (at large $N$). There are $N$ such states. In fact, such a minimal domain wall is just a combination of pairs of excitations discussed above. The minimal domain wall can be created from the ground state by applying a swap operator between spin $i$ in replica $1$ and spin $i$ in replica $2$. Such a swap operator can be decomposed as 
\begin{equation}
    \text{SWAP}_{i,12}= \frac{1}{2}(I +  \bm\sigma_{i}^1 \cdot \bm\sigma^{2}_{i} ).
\end{equation}
where $\bm\sigma_{i}^1 \cdot \bm\sigma^{2}_{i} = \sum_{\alpha}\bm\sigma_{i\alpha}^1  \bm\sigma^{2}_{i\alpha}$. 
Applied to the ground state, this produces a linear combination of the ground state (probability $1/4$) and three pairs of excitations (probability $1/4$ each). Hence, the average energy is $\frac{1}{4} \times 0 + \frac{3}{4} \times 2 \times 12J = 18 J$, just as we found above.

\section{Path integral for the Brownian spin model}
\label{append:spin}

The time evolution operators have the following path integral representation,
\bea
    && [\Tr U(T)]^k [\Tr U(T)^\ast]^k \nn \\
    && = \int d\Omega \exp  \int dt \sum_r \Big[ \sum_{a=L,R;i} \langle \partial_t \Omega_{a,i}^r | \Omega_{a,i}^r \rangle \nn \\
    && - \sum_{i<j} \sum_{\alpha\beta} i J_{ij\alpha\beta} (\sigma_{L,i\alpha}^r \sigma_{L,j\beta}^r - \sigma_{R,i\alpha}^{r} \sigma_{R,j\beta}^{r}) \Big],
\eea
where $\langle \partial_t \Omega_{i}^r | \Omega_i^r \rangle$ is a short-hand notation for spin path integral, and $\Omega$ is a representation for the spin coherent state (such as Euler angular representation). The $L$ and $R$ subindices denote the field in $U(T) = \mathcal T e^{i\int dt H(t)}$ and $U^\ast(T) = \mathcal T e^{i\int dt H^\ast(t)}$ respectively. $\mathcal T$ denotes time ordering. 
More precisely, $\sigma_{L,i\alpha} = \langle \Omega| [{\bm \sigma}_{L,i}]_{\alpha} | \Omega \rangle $ and $\sigma_{R,i\alpha} = \langle \Omega| [\bm \sigma_{R,i}^\ast]_{\alpha} | \Omega \rangle$, where $\bm \sigma$ is the Pauli matrix. 
$r=1,...,k$ is the replica index.  
To get the frame potential, we average over the ensemble by integrating out the Brownian Gaussian variable,
\bea
    && F_{\text{b-spin}}^{(k)} (T) = \int D\Omega \exp \int dt \Big[ \sum_{r,a,i} \langle \partial_t \Omega_{a,i}^r | \Omega_{a,i}^r \rangle \nn \\
    && +  \frac{J}{4N} \sum_{r,s} \sum_{a\ne b} (\sum_{i\alpha} \sigma_{a,i\alpha}^r \sigma_{b,i\alpha}^s)^2\Big] - \frac{9}2 k N J T ,
\eea
where we use that $ \langle \Omega|\sum_{i} \bm \sigma_{a,i} \cdot \bm \sigma_{a,i} | \Omega \rangle = 3N$ for the spin coherent state to get the last term. 
To proceed, we define the Green's function $G^{rs}_{ab}(t) = \frac1N \sum_{i\alpha} \sigma_{a,i\alpha}^r(t) \sigma_{b,i\alpha}^s(t)$, which can be implemented via the following identity
\bea
    1 = \int D F e^{N \int dt \sum_{rs} \sum_{a\ne b} F^{rs}_{ab} (G^{rs}_{ab} - \frac1N \sum_{i\alpha} \sigma_{a,i\alpha}^r \sigma_{b,i\alpha}^s)}. \nn
\eea
Thus, the action becomes 
\bea
    - \frac{I}N 
    &=& \int dt \Big[ \big( \sum_{r,a} \langle \partial_t \Omega_{a}^r | \Omega_{a}^r \rangle - \sum_{rs} \sum_{a\ne b} F_{a b}^{rs} \sum_{\alpha} \sigma_{a,\alpha}^r \sigma_{b,\alpha}^s \big) \nn \\
    && +  \sum_{rs} \sum_{a\ne b} \big( F^{rs}_{ab} G_{ab}^{rs} +  \frac{J}4 (G_{ab}^{rs})^2  \big) \Big] - \frac92 k J T. 
\eea
We can see that the first term gives $N$ copies of the free energy for $2k$ spins, i.e., $r=1,...,k$ and $a,b=L,R$, with the Hamiltonian given by $F^{rs}_{ab}(t)$,
\bea
    && \int D\Omega \exp \int dt \big( \sum_{r,a} \langle \partial_t \Omega_{a}^r | \Omega_{a}^r \rangle - \sum_{rs} \sum_{a\ne b} F_{a b}^{rs} \sum_{\alpha} \sigma_{a,\alpha}^r \sigma_{b,\alpha}^s \big) \nn \\
    && = \Tr[e^{-\int dt H(t)}], 
\eea
and the Hamiltonian reads
\bea
    H(t) = - \sum_{rs} \sum_{a\ne b} F^{rs}_{ab}(t) {\bm \sigma_{a}^r} \cdot {\bm \sigma_{b}^s},
\eea
where we have made a basis transformation for the $R$ spins using $\bm \sigma_{R,2}^s $, which combining with the complex conjugation is equivalent to a time reversal transformation. 
Thus, we obtain the path integral representation of the frame potential for the Brownian spin model with large-$N$ action given by
\bea
    && - \frac{I}N = \log \Tr[e^{-\int dt H(t)}] \nn \\
    && + \int dt \sum_{rs} \sum_{a\ne b} \big( F^{rs}_{ab} G_{ab}^{rs} +  \frac{J}4 (G_{ab}^{rs})^2  \big) -  \frac92 k J T.
\eea

Since the action is quadratic in $G_{ab}^{rs}$, we can integrate out $G_{ab}^{rs}$. 
This amounts to use equation of motion for $G$, i.e., $F_{ab}^{rs} = - \frac{J}2 G_{ab}^{rs}$, to eliminate $G$ field.
It is equivalent to eliminate $F_{ab}^{rs}$, which leads to
\bea
    - \frac{I}{N} &=& \log \Tr e^{-\int dt H(t)} - \frac{J}4 \sum_{rs} \sum_{a\ne b} \int dt (G_{ab}^{rs})^2 - \frac92 k J T. \nn \\
    H(t) &=& \frac{J}2 \sum_{rs} \sum_{a\ne b} G_{ab}^{rs}(t) \bm \sigma_{a}^r \cdot \bm \sigma_{b}^s.
\eea

\subsection{k=1}

We first look at a single replica $k=1$, and then generalize to arbitrary $k$. 
To look for a steady solution, we assume $G_{ab}$ is a constant (here we omit the superscript for the replica index for simplicity).
The Hamiltonian is 
\bea
    H = \frac{J}2 \sum_{a\ne b} G_{ab}\bm\sigma_{a} \cdot \bm\sigma_{b},
\eea
which only involves two spins $a= L, R$. 
The eigenstates include one singlet state and three triplet states. 
We can immediately get the partition function
\bea
    \Tr[e^{-HT}] = e^{\frac{3JT}2 (G_{LR} + G_{RL} ) } + 3 e^{-\frac{JT}2 (G_{LR} + G_{RL}) },
\eea
and the action becomes
\bea
    - \frac{I}N 
    &=& \log [e^{ 3 G_{LR} J T } + 3 e^{-G_{LR} JT }] 
     -    \frac{JT}2 (G_{LR})^2  - \frac92 JT, \nn \\
\eea
where we use the properties $G_{ab} = G_{ba}$ that should be satisfied for a solution.  
The equation of motion is
\bea
     G_{LR} &=& 3 \left( 1 - \frac4{3+ e^{4G_{LR} JT}} \right),
\eea
In the long time limit, the solution is $G_{LR} = 3$, consistent with the assumption of a time independent solution. 
Using this saddle point solution, the frame potential for $k=1$ is 
\bea
    F^{(1)}_\text{b-spin}(T) \approx \big( 1 + 3 e^{-12 J T} \big)^N \approx e^{3N e^{-12 J T}}. 
\eea

\subsection{General k}

Now consider general $k$.
We again assume the solution is time independent to look for steady solutions.
The action can be simplified to
\bea
    - \frac{I}{N} &=& \log \Tr e^{- T H} - \frac{JT}2 \sum_{rs} (G_{LR}^{rs})^2 - \frac92 k J T. \nn \\
    H(t) &=& J \sum_{rs} G_{LR}^{rs} \bm \sigma_{L}^r \cdot \bm \sigma_{R}^s.
\eea
At the long time limit, the first term projects the state to the ground state of $H$.
Let's denote the ground state of $H$ to be $|\Psi\rangle$.
The equation of motion becomes
\bea
    G_{LR}^{rs} = - \langle \Psi | \bm \sigma_{L}^r \cdot \bm \sigma_{R}^s | \Psi \rangle.
\eea
It is clear that for any state $| \Psi \rangle$, $- \langle \Psi | \bm \sigma_{L}^r \cdot \bm \sigma_{R}^s | \Psi \rangle \le 3$. 
The equality is saturated for a spin singlet state between $\bm \sigma_{L}^r$, and $\bm \sigma_{R}^s$, so $G_{LR}^{rs} \le 3$. 

At the long time limit, since the first term projects to the ground state $\log \Tr e^{- T H} = - T \langle \Psi | H | \Psi \rangle $ the action becomes
\bea
    - \frac{I}{N} & \rightarrow & \frac{JT}2 \sum_{rs} (G_{LR}^{rs})^2 - \frac92 k J T.
\eea
We expect the action to be time independent. 
As the second term is linear in the replica, it indicates that the $L$ spin and $R$ spin will form $k$ singlets. 
There are $k!$ different pairing, given by the permutation matrix.
Then it is not hard to verify that at long time limit, the saddle point solution is
\bea
    G_{LR}^{rs} = 3\cdot [P(\pi)]^{rs},
\eea
where $P(\pi)$ is a permutation matrix corresponding to permutation $\pi$. 
For each pair $r$ and $s$ spins, it forms a triplet state.

Indeed, this is consistent with the Hamiltonian analysis. The singlets between the $L$ spin and $R$ spin are precisely the EPR state when we make the transformation back to the original spins. 

To see the finite time correction, we can make the first order perturbation theory by plugging the solution to the finite time action
\bea
    - \frac{I}{N} &=& \log \Tr e^{- 3 JT \sum_{rs} [P(\pi)]^{rs} \bm \sigma_L^r \cdot \bm \sigma_R^s }  - 9 k J T \\
    &=& \log (e^{9JT}+3 e^{-3JT})^k - 9kJT \\
    &\approx&  3 k e^{-12JT}.
\eea

Including the degeneracy from the $k!$ different permutation matrices $P(\pi)$, the frame potential is given by
\bea
    F_\text{b-spin}^{(k)} (T) \approx k! e^{-I} = k! e^{- 3 N k e^{-12 JT}}. 
\eea

\section{Path integral for the Brownian SYK model}
\label{append:syk}


The fermonic path integral for the time evolution operator is given by
\begin{widetext}
\bea
	&& [\Tr U(T)]^k [\Tr U(T)^\ast]^k = \int D\psi \exp \sum_{r} \Big[ i \int dt   \big( \frac12 \psi_{L,i}^r (i\partial_t) \psi_{L,i}^r - J_{ijkl}(t) \psi_{L,i}^r \psi_{L,j}^r \psi_{L,k}^r \psi_{L,l}^r  \big) \nn \\
	&& + i \int dt  \big( \frac12 \psi_{R,i}^r (-i\partial_t) \psi_{R,i}^r + J_{ijkl}(t) \psi_{R,i}^r \psi_{R,j}^r \psi_{R,k}^r \psi_{R,l}^r \big) \Big]
\eea
\end{widetext}
where $a=L,R$ denotes the field in forward $U$ and backward evolution $U^\ast$, respectively, and $r=1,...,k$ denotes different replica. The summation over flavor indices is implicit. 
We can make a redefinition of $\psi_R$, $\psi_R \rightarrow i \psi_R$, so that the path integral becomes
\begin{widetext}
\bea
     [\Tr U(T)]^k [\Tr U(T)^\ast]^k && = \int D\psi \exp \sum_{r}  \int dt   \big( - \frac12 \sum_{a=L,R}\psi_{a,i}^r \partial_t \psi_{a,i}^r - i J_{ijkl}(t) [\psi_{L,i}^r \psi_{L,j}^r \psi_{L,k}^r \psi_{L,l}^r - \psi_{R,i}^r \psi_{R,j}^r \psi_{R,k}^r \psi_{R,l}^r] \big), \nn \\
\eea
And after we average over the disorder distribution, we get
\bea \label{eq:SYK4-path}
    F^{(k)}_\text{bSYK}(T) &=& \int D\psi \exp  \int dt  \Big[ - \frac12 \sum_{a,i,r} \psi_{a,i}^r \partial_t \psi_{a,i}^r + \frac{4 \cdot 3! J}{ N^{3}} \sum_{ijkl} \big[\sum_r ( \psi_{L,i}^r \psi_{L,j}^r \psi_{L,k}^r \psi_{L,l}^r - \psi_{R,i}^r \psi_{R,j}^r \psi_{R,k}^r \psi_{R,l}^r) \big]^2  \Big], \nn \\
\eea
\end{widetext}
where we restore the summation of flavors to be clear.



To derive the large-$N$ action, we introduce the bilocal field $G^{rs}_{ab}(t,t') = \frac1N \sum_{i} \psi_{a,i}^r(t) \psi_{b,i}^s(t')$, where $a,b = L, R$. To implement this equation, we use the following delta function
\bea
    && 1 = \nn \int D \Sigma \exp \frac{N}2 \sum_{rs,ab}\int dt dt' \Sigma^{rs}_{ab} \big(G^{rs}_{ab} - \frac1N\sum_{i} \psi_{a,i}^r \psi_{b,i}^s \big), \nn 
\eea
and we get the effective action~(\ref{eq:bSYK-action}).
The equation of motion is
\bea
    G^{-1} &=& \partial_t + \Sigma, \\
    \Sigma_{ab}^{rs}(t,t') &=& - J \delta(t-t') [2 G_{ab}^{rs}(t,t')]^3. 
\eea

The equation of motion is a complicated matrix equation. But inspired by the invariant state, we observe that when the matrix element for the $G_{LR}$ in the $r$-row and the $s$-column is nonvanishing and all other elements in the $r$-row and the $s$-column are zero, as well as the $G_{aa}$ is diagonal, the equation of motion for this fixed $rs$ decouples from the rest, and is simplified to a two by two matrix equation (i.e. $ \hat G = \left( \ba{cccc} G_{LL}^{rr} & G_{LR}^{rs} \\ G_{RL}^{sr} & G_{RR}^{ss} \ea \right)$. Note that we will sometimes omit the index $rs$ for simplicity). 
Then the equation of motion can be solved with the boundary condition $\psi_{a,i}^r(t + T) = - \psi_{a,i}^r(t + T) $. The solution is given by
\bea \label{eq:solution}
     && \hat G(t_1,t_2) =   \Big[  \frac{\theta(t_{12})}2 f_+(t_{12}) - \frac{\theta(t_{12})}2 f_+(-t_{12}) \Big] 1_{2\times 2} \nn \\
    && \pm \Big[\frac{\theta(t_{12})}2 f_-(t_{12}) - \frac{\theta(t_{12})}2 f_-(-t_{12}) \Big] \sigma_y,  \\
    && \hat \Sigma(t_1, t_2) = \pm \delta(t_{12}) J \sigma_y,
\eea
where $t_{12} = t_1 - t_2$, and
\bea
    f_{\pm}(t) = \frac{ e^{- J t}}{e^{-J T} + 1} \pm \frac{ e^{J t}}{e^{J T} + 1}.
\eea
We have used the Pauli matrix $\sigma_y$ to simply our notation for the solution of the two by two matrix for a fixed $r$ and $s$. 
The plus and minus sign in the front of the $\sigma_y$ means that there two solutions. 


Because this solution is decoupled from the rest, we can calculate the onshell action for a fixed $rs$. The $\log \det$ part can be mapped to a partition function of Majorana fermions with Hamiltonian $ H =\frac12 \hat \Sigma $. Thus we can get the $\log \det$,
\bea
	\frac12 \log \det (\partial_t + \Sigma) &=& \log \int D\chi \exp \left[ - \int dt \chi \Big( \frac12 \partial_t + H \Big) \chi \right] \nn \\
	&=& \log \left( 2\cosh\frac{J T}2 \right).
\eea
Other terms can be evaluated directly, then the onshell action for the $rs$ component gives
\bea \label{eq:slowestfermion}
    - \frac{I_{\text{wh}}(T)}{N} = \log \left( 2\cosh\frac{J T}2 \right) - \frac{JT}2,
\eea
where the subscript $\text{wh}$ indicates that it is like a wormhole solution connecting $r$-th and $s$-th replicas.

On the other hand, there is an obvious trivial solution given by $\hat G = [\sgn(t_{12})/2] \cdot 1_{2\times 2} $, $\hat \Sigma = 0$, with onshell action given by  
\bea
- \frac{I_{0}(T)}{N} = \log2 - \frac{JT}8, 
\eea
where 
the subscript $0$ indicates that it comes from a trivial solution. 

Restricted to the $r$-th and $s$-th replicas, when there is a nonvanishing correlation between them, the onshell action tends to zero at long time limit. 
On the other hand the trivial solution tends to infinity at the long time limit, indicating that they do not contribute to the frame potential: it is exponentially suppressed. 
While at time zero, both solutions leads to the dimension of Hilbert space, $2^N$. 
This number is because each replica contains $N$ Majorana fermoins.

The above solutions are restricted to a two by two matrix, i.e., the $r$-th and $s$-th replicas, and the remaining step is to count how many different solutions can exist among the $k$ replicas. 
The result is given by
\bea \label{eq:saddle}
	F_{\text{bSYK}}^{(k)}(T) 
	&=&  \sum_{m=0}^k 2^m m! \left(\ba{cccc} k \\m \ea\right)^2 e^{-(k-m)I_0(T)} e^{- m I_\text{wh}(T)}, \nn \\
\eea
where $m$ indicates the number of nontrivial paired solutions and $k-m$ indicates the number of trivial solutions in the $k$ possibilities. 
The prefactor is the degeneracy. 
It comes from picking $m$ elements out of $k$ possibilities for the row and the column (so there is a square), and shuffling the result, while the $2^m$ factor accounts for the two solutions corresponding to different sign choices in~(\ref{eq:solution}). 
At time zero, the frame potential should give the dimension of Hilbert space of $2k$ Brownian SYK systems. 
It seems that our result is much greater than the dimension of Hilbert space because the contribution from nontrivial solutions are included. 
But as we will see in the following, those nontrivial solutions should not be included at time zero. 

At the long time limit, the trivial solution will lead to an exponential suppression with exponent proportional to $N$, i.e.,
\bea
    e^{-(k-m)I_0(T)} = e^{(k-m) N (\log 2 - \frac{JT}8)}, 
\eea
for $m<k$. 
Thus, the dominant solution is given by $m=k$ with maximal pairs in the $2k$ different replicas, 
\bea \label{eq:frame-potential}
    F_{\text{bSYK}}^{(k)}(T) &=& 2^k k! e^{k N \log (1 + e^{- J T})} \\
    &\approx& 2^k k! (1+ k N e^{- J T}).
\eea
Regarding the prefactor, while $2^k$ is given by the sign choice, $k!$ corresponds nicely to the permutation. 


\section{$1/N$ correction in the Brownian SYK model}
\label{append:correction}

We consider the small fluctuations around the saddle point solution,
\bea
    G(t_1,t_2) &=& \bar G(t_1-t_2) + \frac1{\sqrt N} g(t_1, t_2), \\
    \Sigma(t_1,t_2) &=& \delta(t_{12}) \left(\bar \Sigma + \frac1{\sqrt N} \sigma\Big( \frac{t_1+t_2}2 \Big) \right).
\eea
Here we use $\bar G$ and $\bar \Sigma$ to denote the solution, and $g$ and $\sigma$ denote the fluctuation. The factor $\frac1{\sqrt N}$ is a proper rescaling for the large $N$ theory.

The potential $\frac{J \delta(t-t')}{16} c_{ab} (2  G_{ab}^{rs})^4$ indicates that only when $\bar G^{rs}_{ab} \ne 0$ can one gets a nonvanishing contribution in the quadratic order. 
Thus, we focus on these fluctuations for each fixed $rs$ with nontrivial solutions. 
We omit the indices $rs$ for simplicity. 
Other fluctuations will not have a contribution at quadratic level and will also force $\sigma$ to be zero.

We expand each terms in the action up to quadratic orders in the fluctuations. 
The $\log \det$ leads to
\bea \label{eq:sigma1}
    && - \frac14 \frac1{T^2} \sum_{\Omega,\omega} \Tr[\bar G(\Omega + \omega) \sigma(\omega) \bar G(\Omega) \omega(-\omega)] \nn \\
    && = \frac1{T} \sum_\omega \frac12 \frac{\tanh \frac{JT}2}{\omega^2 + 4J^2} \nn \\
    && \times \left(\ba{cccc} \sigma_{11}(-\omega) & \sigma_{12}(-\omega) \ea \right) \left( \ba{cccc} 2J & \omega \\ - \omega & 2J \ea \right) \left(\ba{cccc} \sigma_{11}(\omega) \\ \sigma_{12}(\omega) \ea \right), 
\eea
where we have used the basis transformation $\sigma_{LL/RR} = \sigma_{11} \pm \tilde \sigma_{11}$, $\sigma_{LR/RL} = \sigma_{12} \pm \tilde \sigma_{12}$. 
The Fourier transformation is defined through
\bea
    \sigma_{ab}(\omega_n) &=& \int_0^T dt \sigma_{ab}(t) e^{i \omega_n t}, \\ \sigma_{ab}(t) &=& \frac1T \sum_n \sigma_{ab}(\omega_n) e^{-i \omega_n t} ,
\eea
where $\omega_n = \frac{2\pi n}{T}$ is the Matsubara frequency for bosons. 
We denote $\omega_n$ as $\omega$, and $\sum_n$ as $\sum_\omega$ for simplicity. 

The term $\int dt dt' \frac12 \Sigma_{ab}^{rs} G_{ab}^{rs} $ leads to
\bea
    \int dt \frac12 \Big[ (g_{LL}(t) + g_{RR}(t)) \sigma_{11}(t) + (g_{LR}(t) + g_{RL}(t)) \sigma_{12}(t) \nn \\
    + (g_{LL}(t) - g_{RR}(t)) \tilde\sigma_{11}(t) + (g_{LR}(t) - g_{RL}(t)) \tilde\sigma_{12}(t) \Big], \nn
\eea
where because the $\delta$ function resulted from Brownianness, only $g(t,t)$ is relevant, and we define $g_{ab}(t) \equiv g_{ab}(t,t)$. Because $\tilde\sigma_{11}$ and $\tilde\sigma_{12}$ are zero modes, they will serve as Lagrange multipliers to force $g_{LL} = g_{RR}$ and $g_{LR} = g_{RL}$. 
Then the above expression reduces to
\bea \label{eq:sigma2}
    \int dt ( g_{LL}(t) \sigma_{11}(t) + g_{LR}(t) \sigma_{12}(t) ).
\eea

Combining (\ref{eq:sigma1}) and (\ref{eq:sigma2}), and integrating out $\sigma_{11}$ and $\sigma_{12}$, we get
\bea
    && - \frac1T \sum_\omega \frac12 \coth \frac{JT}2 \times \nn \\
    && \left(\ba{cccc} g_{LL}(-\omega) & g_{LR}(-\omega) \ea \right) \left( \ba{cccc} 2J & -\omega \\ \omega & 2J \ea \right) \left(\ba{cccc} g_{LL}(\omega) \\ g_{LR}(\omega) \ea \right).
\eea

And finally, the term $ \int dt dt'\frac{J \delta(t-t')}{16} c_{ab} (2  G_{ab}^{rs})^4$ gives rise to $-6 J \frac1T \sum_\omega g_{LR}(-\omega) g_{LR}(\omega)$. Putting everything together, we have the final result,
\bea \label{eq:vacuum}
    && -\delta I = \frac1T \sum_\omega - \frac12 \times \nn \\
    && \hat g(-\omega) \left( \ba{cccc} 2J \coth \frac{JT}2 & -\omega \coth \frac{JT}2 \\
    \omega \coth \frac{JT}2 & 2J \coth \frac{JT}2 + 12 J \ea \right) \hat g(\omega), 
\eea
where $\hat g = (g_{LL}, g_{LR})^T$. 
The free energy of this free boson can be evaluated as
\bea
    F &=& \frac{1}T \sum_n \frac12 \log(\omega_n^2 + E^2) = \frac{1}T \log (1- e^{-\beta E}), 
\eea
where $E = 2J \sqrt{6 \tanh \frac{JT}2 - 1}$ is the mass of the real boson. 
There is a critical $T^*$, i.e., $ \tanh \frac{JT^*}2 = \frac16 $, after which these boson becomes stable. Thus, when $T<T^*$, we should not include these saddle point contributions as we have discussed before. 

For a fixed permutation solution, we have $k$ different $rs$, which leads to $k$ bosons. Now we have same degeneracy coming from $2^k k!$. For the late times, the free energy for each of the degenerate solution including the quadratic fluctuations reads 
\bea
    - F T = k N \log (1+ e^{- J T}) - k \log(1 - e^{- E T})\Big( 1 + \mathcal{O}(\frac1N) \Big), \nn 
\eea
Inside the bracket of the second term, $1$ is from the quadratic fluctuation~(\ref{eq:vacuum}), and $\mathcal{O}(\frac1N)$ is given by the vacuum bubble.
Because we are interested in the long time behavior, we expect a renormalization flow of the mass $E$ due to high order corrections gives an appropriate estimate of the collective decaying mode. 
The large $N$ structure indicates that correction is of order $\mathcal{O}(\frac1N)$, namely, we expect the free energy is given by
\bea
    - F T = k N \log (1+ e^{- J T}) - k \log(1 - e^{- \big(E + \mathcal{O}( \frac{1}N ) \big) T}). \nn \\ 
\eea

\section{Effective Hamiltonian for the Brownian SYK model}
\label{append:syk-hamiltonian}

In (\ref{eq:SYK4-path}), the boundary condition is for any field configurations, so the above path integral can be regarded as a partition function 
\bea
    \overline{[\Tr U(T)]^k [\Tr U(T)^\ast]^k} = \Tr[e^{- H T }],
\eea
with the Hamiltonian given by,
\bea
    && H = \frac{4 \cdot 3! J}{ N^{3}} \\
    && \sum_{ijkl} \big[\sum_r ( \psi_{L,i}^r \psi_{L,j}^r \psi_{L,k}^r \psi_{L,l}^r - \psi_{R,i}^r \psi_{R,j}^r \psi_{R,k}^r \psi_{R,l}^r) \Big].
\eea
The ground state of this Hamiltonian is given by the tensor product of EPR state, each of which is two EPR states of two contours $r$ and $s$
\bea
    |\text{GS} \rangle = \otimes_{r,s} |\infty \rangle_{r,s}, \quad 
    (\psi^{r}_j \pm i \psi^{s}_j )| \infty \rangle_{r, s} = 0, \quad \forall j.
\eea
Then the number of such ground states is given by 
\bea
    N(\text{GS}) = 2^k k!.
\eea

The elementary excitation (eigenstate) is given by $\psi_i^r|\text{GS} \rangle $ for any flavor $i$ and replica $r$. The energy of this excitation is
\bea
    H \psi_i^r|\text{GS} \rangle = J \psi_i^r|\text{GS} \rangle.
\eea
The number of the elementary excitation is given by $kN$ for each degenerate ground state, so we expect the frame potential is given by
\bea
    F_{\text{bSYK}}^{(k)}(T) = 2^k k! (1 + k N e^{-JT}),
\eea
which exactly reproduces the saddle point calculation~(\ref{eq:frame-potential}).

\subsection{Quadratic SYK model}
Let's consider the quadratic SYK model given by
\bea
	H(t) =  \sum_{i<j} J_{ij}(t) \psi_{i} \psi_{j},
\eea
where $\psi_{j}$, $j=1,...,N$, $\{ \psi_i, \psi_j \} = \delta_{ij}$ are Majorana fermions. $J_{ij}(t)$ is a Brownian Gaussian variable with mean zero and variance
\bea
	\avg{[J_{ij}(t)J_{i'j'}(0)]} = \delta_{i i'} \delta_{j j'}  \frac{4 J}{N} \delta(t).
\eea
Paralleled to the regular SYK model, we arrive at the path integral, 
\begin{widetext}
\bea
    [\Tr U(T)]^k [\Tr U(T)^\ast]^k &=& \int D\psi \exp \sum_{r} \Big[ \int dt   \big( - \frac12 \sum_i \psi_{L,i}^r \partial_t \psi_{L,i}^r - i \sum_{ij} J_{ij}(t) (i \psi_{L,i}^r \psi_{L,j}^r ) \big) \nn \\
	&& + \int dt  \big( - \frac12 \sum_i \psi_{R,i}^r \partial_t \psi_{R,i}^r - i \sum_{ij} J_{ij}(t) (i \psi_{R,i}^r \psi_{R,j}^r ) \big) \Big].
\eea
\end{widetext}
Owing to the similarity between $L$ and $R$ fields after the redefinition, we can include these two indices into $r$, such that $r = 1, ..., 2k$,
\bea
    && [\Tr U(T)]^k [\Tr U(T)^\ast]^k = \\
    && \int D\psi \exp \sum_{r=1}^{2k}  \int dt  \Big[ - \frac12 \sum_i \psi_{i}^r \partial_t \psi_{i}^r - i \sum_{ij} J_{ij}(t) (i \psi_{i}^r \psi_{j}^r ) \Big]. \nn \\
\eea
Then apparently, the action has a $O(2k)$ symmetry. Integrating out the Brownian variable, we have 
\bea
    && \avg{[\Tr U(T)]^k [\Tr U(T)^\ast]^k} = \\
    && \int D\psi \exp  \int dt  \Big[ - \frac12 \sum_{i,r} \psi_{i}^r \partial_t \psi_{i}^r + \frac{J}{N} \sum_{i<j}  (\sum_r \psi_{i}^r \psi_{j}^r)^2  \Big]. \nn 
\eea
where the Hamiltonian can be read out as
\bea
    H = \frac{J}{N} \sum_{i<j}  \big(\sum_r i \psi_{i}^r \psi_{j}^r \big)^2.
\eea
The ground state of this Hamiltonian is given by the tensor product of EPR state, each of which is two EPR states of two contours $r$ and $s$
\bea
    |\text{GS} \rangle = \bigotimes_{r,s} |\infty \rangle_{r,s}, \quad 
    (\psi^{r}_j \pm i \psi^{s}_j )| \infty \rangle_{r, s} = 0, \quad \forall j. \nn \\
\eea
Then the number of such ground states is given by 
\bea
    N(\text{GS}) = \frac{(2k)!}{k!} \gg 2^k k!,
\eea
which indicates that the SYK$_2$ model is not able to reach $k$ design.

\section{Non-Hermitian Perturbation Theory}
\label{append:nonhermitianperttheory}

Here we derive the perturbative decay rates Eqs. \eqref{eq:decayratenondegen} and \eqref{eq:decayratedegen}, starting from the effective Hamiltonian Eq. \eqref{eq:effhamrmt}, in the limit of small $g \ll 1$.
We write the effective Hamiltonian in the form
\begin{equation}
    H_{\mathrm{eff}} = H + g V
\end{equation}
where
\begin{align}
    H &= \sum_r H^r - H^{\overline{r}} \nonumber \\
    V &= - \frac{i}{2} \sum_{\alpha} \left( \sum_{r} [O^r_{\alpha} - (O_\alpha^{\bar{r}})^*] \right)^2
\end{align}
are the bare Hamiltonian and non-Hermitian perturbation, respectively.

\subsection{$k = 1$}

Consider first the simplest case $k = 1$ and a single perturbing operator $O_{\alpha} = O$, where the bare Hamiltonian and perturbation are
\begin{align}
    H &= H^1 - H^{\overline{1}} \nonumber \\
    V &= - \frac{i}{2} \left( O^1 - O^{\overline{1}} \right)^2.
\end{align}
The bare Hamiltonian $H$ has eigenstates $\ket{\psi^{(0)}_{nm}} \equiv \ket{n,m}$ with bare eigenenergies $E^{(0)}_{nm} = E_n - E_m$. We assume the energies $E_n$ are nondegenerate, but the replicated eigenenergies $E_{nm}$ are clearly degenerate whenever $n = m$. The decay rate for each eigenstate $\ket{\psi_{nm}}$ appears as an imaginary component in the perturbed eigenenergies $E_{nm}$. We can compute this imaginary component explicitly using first-order perturbation theory.

First consider the case $n \neq m$. Similar to standard perturbation theory, we expand the perturbed eigenstate and eigenenergies in powers of the small parameter $g$:
\begin{align}
    \ket{\psi_{nm}} &= \ket{\psi_{nm}^{(0)}} + g \ket{\psi^{(1)}_{nm}} + \mathcal{O}(g^2) \nonumber \\
    E_{nm} &= E_{nm}^{(0)} + g E_{nm}^{(1)} + \mathcal{O}(g^2)
\end{align}
To lowest order in $g$, the Schr{\"o}dinger equation is just the bare eigenvalue equation $(H - E_{nm}^{(0)}) \ket{\psi^{(0)}_{nm}} = 0$. At first order in $g$, the Schr{\"o}dinger equation yields
\begin{equation}
    g H \ket{\psi^{(1)}_{nm}} + g V \ket{\psi^{(0)}_{nm}} = g E^{(0)}_{nm} \ket{\psi_{nm}^{(1)}} + g E^{(1)}_{nm} \ket{\psi_{nm}^{(0)}}
\end{equation}
Taking the inner product with $\bra{\psi_{nm}^{(0)}}$ and using the fact that the bare Hamiltonian is Hermitian $\adj{H} = H$ yields the first result Eq. \eqref{eq:decayratenondegen}, where $\delta E = g E_{nm}^{(1)} \propto - i g$.

Next consider the case $n = m$. This is a $D$-dimensional subspace spanned by the vectors $\ket{n,n}$ for $n = 1,\ldots,D$. As discussed in the main text, we first identify the dark state $\ket{d} = \frac1{\sqrt D} \sum_n \ket{n,n}_{1\overline{1}}$ within this subspace and remove it. We then orthogonalize the remaining vectors so that they span the $D-1$-dimensional subspace orthogonal to $\ket{d}$. The states in this orthogonal subspace are random vectors.
We analyze the spectrum in the subspace in Appendix~\ref{append:perturb-matrix}.


\subsection{$k > 1$}

The argument proceeds similarly for $k > 1$, but now there are many more ways for the eigenenergies
\begin{equation}
    E_{n_1 m_{\overline{1}} n_2 m_{\overline{2}} \ldots n_{k} m_{\overline{k}}} = \sum_{r} E_{n_r} - \sum_{\overline{r}} E_{m_{\overline{r}}}
\end{equation}
to be degenerate. In particular, we can always find $k!$ dark states. One of these is simply
\begin{equation}
    \bigotimes_r \ket{n,n}_{r\overline{r}},
\end{equation}
which is the $2k$-replica generalization of Eq. \eqref{eq:darkstate}. The remaining dark states can be found by permuting the forward replicas $r$ (or the backward replicas $\overline{r}$) amongst themselves, giving $k!$ dark states total. As mentioned in the main text, these dark states are not all orthogonal, but have overlaps that are suppressed by powers of the Hilbert space dimension $D$. Nevertheless, these states still span a space of dimension $k!$ provided $k<D+1$. It is very interesting to consider what happens for larger $k$, especially in the context of computing replica limits, but we leave these technical issues for future work.

Here we analyze the decay rate using the first-order perturbation theory.
While it is known that one can construct an set of exact excited states using the eigenstate from $k=1$~\cite{banchi2017driven}, we argue that in the perturbation theory, the smallest gap is indeed given by individual excitation from $k=1$. 
As before, the perturbation is given by
\bea
    V &=& \frac{-i}2 \left( \sum_{i=1}^k (O^i - O^{\bar i}) \right)^2 \\
    &=&  - i \left( k - \sum_i O^i O^{\bar i} + \sum_{i\ne j} [ O^i O^j + O^{\bar i} O^{\bar j}] -  \sum_{i \ne  j} O^i O^{\bar j}   \right), \nn 
    \label{eq:perturbation-V} \\
\eea 
where in the second line, we separate the diagonal part $i=j$ and the off-diagonal part $i \ne j$. 
Notice here we take $O$ as a Pauli-Z matrix with $O^\ast = O^T = O$ and $O^2 = 1$. 

For a general eigenstate state of the unperturbed Hamiltonian, such as, $|n_1...n_k, m_{\bar 1}...m_{\bar k}\rangle$, we have $(k!)^2$ fold degeneracy by permuting the $r$ and the $\bar r$ replicas independently, i.e., $|n_{\pi(1)}...n_{\pi(k)}, m_{ \pi'(\bar 1)}...m_{\pi'( \bar k)}\rangle$. 

Thus we should look at the matrix element of $V$ in this degenerate subspace. 
The diagonal elements are 
\bea
    && \langle n_1...n_k, m_{\bar 1}...m_{\bar k} | V | n_1...n_k, m_{\bar 1}...m_{\bar k} \rangle \nn \\
    && = -i \left( k - \sum_{i=1}^k \langle n_i | O | n_i \rangle \langle m_{\bar i} | O | m_{\bar i} \rangle \right). 
\eea
Here we neglect the off-diagonal ones because they will cancel on average. 
Following the same estimate from the main text, $\langle n | O | n \rangle \sim 1/\sqrt{D}$, the diagonal elements are approximate by 
\bea \label{eq:diag}
    -i k \left( 1- \mathcal{O} (1/D) \right). 
\eea
We see that the diagonal it is enhanced by a factor of $k$. 

Let's consider the effect from the off-diagonal elements. 
Since we have the coupling between at most two replicas, the perturbation $V$ can only connect two degenerate states related by a transposition.
For transposition $(n_i, n_j)$, 
the operator that connects these two states is $O^i O^j$, so we have the following matrix element
\bea
    &&\langle n_1...n_i...n_j...n_k, m_{\bar 1} ... m_{\bar k} | V | n_1...n_j...n_i...n_k, m_{\bar 1} ... m_{\bar k} \rangle \nn \\
    &=& -i |\langle n_i | O | n_j \rangle|^2.
\eea
And this gives us the transition amplitude which we analyze in Appendix~\ref{append:perturb-matrix}. 
The upshot is that the variance of this matrix element is $2/D^2$.

The perturbation matrix has dimension $(k!)^2 \times (k!)^2$. 
Its diagonal element is approximated by~(\ref{eq:diag}).
Its off-diagonal element is nonzero only when two states can be connected by a transposition. 
As there are $2 \cdot \left(\ba{cccc} k \\ 2 \ea\right) \sim k^2$ transpositions. 
The factor of two is because we can have transpositions from $r$ replicas or $\bar r$ replicas. 
This is a sparse random matrix with approximate $k^2$ nonzero elements in a row with a total $(k!)^2$ elements.  
We approximate this sparsity by introducing a probability $p= \frac1{(k!)^2} \left(\ba{cccc} k \\ 2 \ea\right)  $ such that we have $1-p$ probability to have a vanishing entry and $p$ probability to have an entry with variance $2/D^2$. 
It turns out that when $k \gg 1$, the eigenvalues of this matrix will satisfy the Wigner semicircle law~\cite{rodgers1988density}.
The radius of the semicircle is given by
\bea
    \sqrt{(k!)^2 \cdot p \cdot \frac{2}{D^2}} \sim \frac{k}{D}.
\eea
It seems that the correction is enhanced by a factor of $k$.
But the diagonal element is also enhanced by a factor of $k$, so the off-diagonal element only leads to a $1/D$ correction to the diagonal element~(\ref{eq:diag}). 
Thus the decay rate for a generic state is $kg(1- \mathcal{O}(1/D))$. 

Like the situation in $k=1$, we should pay special attention to the degenerate subspace spanned by $|n_1...n_k, n_1...n_k \rangle$.
The matrix element is
\bea \label{eq:matrix_k}
    && \langle m_1...m_k,m_1...m_k | V | n_1...n_k,n_1...n_k \rangle \nn \\
    &=& -i \left(k \prod_{i=1}^k \delta_{m_i,n_i} - \sum_{i=1}^k |\langle m_i |O| n_i \rangle |^2 \prod_{j\ne i} \delta_{m_j,n_j} \right). \nn \\
\eea
It is straightforward to check that off diagonal parts vanish in~(\ref{eq:perturbation-V}).

Denote the nontrivial matrix appeared in~(\ref{eq:matrix_k}) by
\bea
M_{mn} = |\langle m |O| n \rangle |^2,
\eea
the information of eigenvalues of $M$ determines the full spectrum in~(\ref{eq:matrix_k}). 
To see that, denote the eigenvectors of $M$ by $|E_i \rangle , i= 0,...,D-1$ with eigenvalues $E_i$.
There is a special state given by
\bea
    |E_0 \rangle \equiv |d \rangle = \frac1{\sqrt D} \sum_n |n,n\rangle. 
\eea
It has eigenvalue one $E_0=1$. 
Apart from $E_0$, we assume $E_1>...>E_{D-1}$. 
The other $D-1$ states are random vectors and will be analyzed in detail in Appendix~\ref{append:perturb-matrix}. 

All eigenstates of~(\ref{eq:matrix_k}) can be constructed as
\bea
    |E_{i_1},...,E_{i_k} \rangle = \bigotimes_{j=1}^k |E_i \rangle_j,
\eea
with eigenvalues 
\bea
    -i \left(k - \sum_{j=1}^k E_{i_j} \right).
\eea
Then we see that among the $D^k$ eigenstate of this matrix, there is exactly one zero mode $|E_0,...,E_0 \rangle$. 
It is now clear that the first excited states are
\bea 
    |\Psi_j \rangle &=& |E_0,...,E_1,...,E_0 \rangle \\
    &=& \left( \bigotimes_{i=1}^{j-1} |E_0 \rangle_i \right) \otimes |E_1\rangle_j \otimes \left(\bigotimes_{i=j+1}^{k} |E_0 \rangle_i \right),
\eea
for $j=1,...,k$, namely, there are $k$-fold degeneracy for first excited state, and the eigenvalue is 
\bea
    -i \left(k - (k-1) E_0 - E_1 \right) = -i (1-E_1).
\eea
Because the eigenvalue of $M$ satisfies Wigner semicircle law with radius $\sim 1/\sqrt{D}$ (Appendix~\ref{append:perturb-matrix}), $E_1 < 1/\sqrt{D}$. 
The minimal decay rate is given by 
\bea
    -ig(1 - \mathcal{O}(1/\sqrt{D})),
\eea
which is the same as $k=1$. 

Due to permutation symmetry, we actually have $k!$ such subspace spanned by $|n_1...n_k, n_{\pi(1)}...n_{\pi(k)} \rangle$. 
We expect the matrix element between different subspace is similar to the analysis before except now we have a much bigger matrix, i.e., it is enlarged by a factor of $D^k$.
As a consequence, the correction will be highly suppressed by $D^{-k/2}$, and it is safe to neglect this correction. 

Therefore, for small $g$ and arbitary $k\le D$, we expect that there are $k!$ dark states and $k\cdot k!$ degenerate states with the minimal decay rate $g(1- \mathcal{O}(1/\sqrt{D}))$. 
We have also numerically checked that the averaged minimal gap is independent of $k$ for $k=1,2$.

\section{Eigenvalues of the perturbation matrix} \label{append:perturb-matrix}

In this section, we analyze the eigenvalue of the matrix in the degenerate subspace in the first-order perturbation theory for $k=1$. 
There is a dark state with zero decaying rate, apart from this, we are interested in the decaying rate of other states in the degenerate subspace $|nn\rangle$, $n=1,...,D$. 

The nontrivial matrix is~(\ref{eq:perturb-matrix}) and also given as follows,
\begin{equation} 
    M_{nm} = |\langle n | O |m \rangle|^2, 
\end{equation}
where $O$ is taken to be a Pauli-Z matrix, and $|n\rangle$ is an eigenvector of a GOE matrix. We are interested in eigenvalues of $M$.

Because the constraint for the eigenvector is that they are real and satisfy the normalization, an eigenvector of a GOE random matrix has the distribution~\cite{brody1981random}
\bea
    \rho(x_1,..,x_D) = \pi^{-D/2} \Gamma\Big( \frac{D}2 \Big) \delta\Big(\sum_{i=1}^D x_i^2 - 1 \Big),
\eea
where $x_i$ is each component of the eigenvector. It can be understood as the projection of an eigenvector onto $n$ orthonormal axes. 
Because eigenvectors are orthogonal, it can be also understood as projection of all eigenvectors onto a fixed normal axis. 
These two quantities are the same in orthogonal ensembles. 
We can integrate out $D-1$ components to get 
\bea
    \rho(x) = \frac{\Gamma(\frac{D}2)}{\pi^{1/2} \Gamma( \frac{D-2}2) } (1-x^2) ^{\frac{D-3}2}.
\eea
This can be interpreted as projection of an eigenvector to a fixed unit axis. In the large $D$ limit, it is easy to see that $x \sim D^{-1/2}$, and the distribution tends to
\bea
     \rho(x) \rightarrow \left( \frac{D}{2\pi} \right)^{1/2} e^{- \frac{D}2 x^2}.
\eea

Now consider the transition amplitude of the Hermitian operator $O$, $O_{nm} = \langle n | O | m \rangle$. 
To estimate this quantity at large $D$ limit, we first imagine $m$ is fixed and then allow $m$ to move orthogonally. 
When the $m$-th eigenvector is fixed, the transition amplitude is equal to the projection of $n$-th eigenvector to a fixed axis $O|m\rangle$. 
An essential difference is that this axis in general is not necessary a unit axis. 
The norm is 
\bea
    \sigma_m^2 = \langle m | O^\dag O |m \rangle.
\eea
So simply using change of random variables, the distribution of projection is
\bea    
    \rho_{O_{nm}}(x;\sigma_m) = \frac1{\sigma_m} \rho\left( \frac{x}{\sigma_m} \right),
\eea
where the second argument implies the distribution depends on the norm of the axis. 
Now we allow the $m$-th eigenvector to move orthogonally as well, so $\sigma_m$ will change accordingly. 
Assume $\sigma_m$ satisfies a distribution $\rho_{\sigma_m}(x)$, if we know this distribution, then we have~\cite{brody1981random} 
\bea
    \rho_{O_{nm}}(x) = \int dz \rho_{O_{nm}}(x;z) \rho_{\sigma_m}(z). 
\eea

For our purpose, the norm is actually fixed due to the factor that $O^\dag O = 1$ for Pauli-Z operator. Thus $\sigma_m = 1$, and in the large $D$ limit,
\bea
    \rho_{O_{nm}}(x) = \left( \frac{D}{2\pi} \right)^{1/2} e^{- \frac{D}2 x^2}.
\eea
Consistent with our intuition, the distribution does not depend on $n$ and $m$. We will also need 
\bea \label{eq:norm-distribution}
    \rho_{|O_{nm}|^2}(x) = \left( \frac{D}{2\pi x} \right)^{1/2} e^{- \frac{D}2 x}.
\eea

From the above discussion, we know that the matrix element $M_{nm}$ satisfies identical independent distribution~(\ref{eq:norm-distribution}). 
Thus it is a Wigner matrix. We expect its eigenvalues will satisfy Wigner's semicircle law. 
According to this distribution, the mean is $1/D$ and the variance is $2/D^2$, Wigner's semicircle law predicts
\bea
    \rho(E) = \frac{\sqrt{D}}{2\pi} \sqrt{2- \frac{D}4 
    \left(E- \frac1D \right)^2},
\eea
which agrees well with a numerical calculation of the random matrix, see Fig.~\ref{fig:semicircle}

\section{Strong complexity, $k$-designs, frame potential}
\label{append:strongcomplexity}

Our calculations of the $k$-th frame potential immediately tell us about the growth of circuit complexity in these models. Here we review the connection between $k$-designs, the $k$-th frame potential, and circuit complexity, following the arguments in \cite{brandao2021models,roberts2017chaos}. These connections can be motivated by a simple observation: it is very difficult to distinguish a $k$-design $\mathcal{U}$ from the completely depolarizing channel $\mathcal{D}$ that sends every input state $\rho_0$ to the infinite-temperature state:
\begin{equation}
    \mathcal{D}(\rho_0) = \id / \tr{[\rho_0]}.
\end{equation}
From the perspective of local observables, both of these channels produce maximally mixed states, so we cannot tell them apart unless we can prepare and measure very complicated states with substantial many-body correlations.

\begin{figure}
    \centering
    \includegraphics[width=0.9\columnwidth]{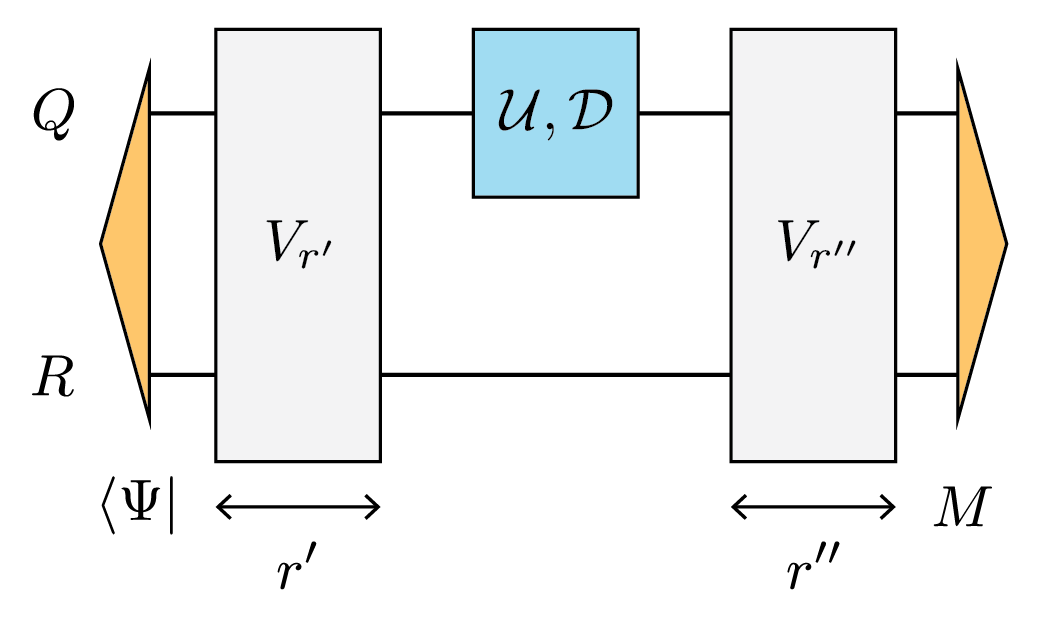}
    \caption{Distinguishing a unitary channel $\mathcal{U}$ from the depolarizing channel $\mathcal{D}$. To distinguish $\mathcal{U}$ from $\mathcal{D}$, Bob is allowed to use an ancilla system $R$, along with arbitrary pre- and post-preparation circuits of depth $r'+r'' = r$. Bob cannot reliably distinguish a $k$-design $\mathcal{U}$ from $\mathcal{D}$ unless he has circuits of depth at least $r \sim k N / \log N$.}
    \label{fig:diamondcircuit}
\end{figure}

We can phrase this observation in concrete terms by posing a simple adversarial game between two players, Alice and Bob. In this game, Alice constructs either an $\epsilon$-approximate $k$-design $\mathcal{U}$ on $N$ qubits or the completely depolarizing channel $\mathcal{D}$ and hands this channel to Bob; Bob's task is to distinguish which channel $\mathcal{U},\mathcal{D}$ has been given to him using only a single use of the channel. To complete this task, Bob is allowed to use an $N$-qubit reference system $R$ as well as pre- and post-preparation circuits of total depth $r = r' + r''$ (see Fig. \ref{fig:diamondcircuit}). Using these tools, Bob looks for a `bias' signal
\begin{equation}
    S \equiv \beta^{\sharp}(r,\mathcal{U}) \equiv \magn{\tr{\left[ M \left( \mathcal{U} \otimes \mathcal{I} - \mathcal{D} \otimes \mathcal{I} \right) \left( \ket{\Psi}\bra{\Psi} \right) \right]}}
\end{equation}
which indicates whether the channel is distinct from the depolarizing channel, where $\mathcal{I}$ is the identity channel acting on the reference $R$. Bob attempts to maximize this bias signal by searching over all possible initial states $\ket{\Psi}$ and projective measurements $M$, which are prepared by the pre- and post-processing circuits of depth $r',r''$, respectively. (Because of the cyclicity of the trace, we actually don't need to search over initial states $\ket{\Psi}$ and measurements $M$ separately. It suffices to fix $\ket{\Psi}$ as a maximally-entangled TFD state between $Q,R$ and consider a search over all measurements $M$ that can be prepared by circuits of depth $r$ or less.) The gist of the argument is that the probability $\mathrm{Pr}[ S \geq \tau ]$ for Bob to find a substantial bias is extraordinarily small if he is only allowed to use short circuits of depth $r \lesssim k N / \log N$ or less. In this sense, we say that a $k$-design $\mathcal{U}$ has circuit complexity $r$ that grows linearly with $k$.


We now provide a more rigorous proof of these arguments. Following Ref. \cite{brandao2021models}, we say that a channel $\mathcal{U}$ has strong $\delta$-unitary complexity at most $r$ if the optimal bias signal is sufficiently close to its optimal value:
\begin{equation}
    S = \beta^{\sharp}(r,\mathcal{U}) \geq 1 - 1/D^2 - \delta.
\end{equation}
In the following, we prove that every $\epsilon$-approximate $k$-design $\mathcal{U}$ has strong $\delta$-unitary complexity at most $r \sim k N / \log N$.

To show this we utilize Markov's inequality,
\begin{equation}
    \mathrm{Pr}[S \geq \tau ] = \mathrm{Pr}[S^{2k} \geq \tau^{2k}] \leq \tau^{-2k} \mathbb{E} [ S^{2k} ]
\end{equation}
which bounds the probability of obtaining a substantial bias signal $S \geq \tau = 1 - 1/D^2 - \delta$ in terms of the $2k$-th moments of the bias.
If the channel $\mathcal{U}$ is an $\epsilon$-approximate $k$-design, then the right-hand side can be bounded by
\begin{align}
    \label{eq:markov2kdesignbound}
    \tau^{-2k} \mathbb{E} [ S^{2k} ] &\equiv \tau^{-2k} \mathbb{E} \left[ \tr{\left[ M \left( \mathcal{U} \otimes \mathcal{I} - \mathcal{D} \otimes \mathcal{I} \right) \left( \ket{\Psi}\bra{\Psi} \right) \right]^{2k} } \right] \nonumber \\
    &\leq \frac{((2k)!)^2}{D^k \tau^{2k} } \left(C_{2k} + \frac{\epsilon}{(2k)! D^{3k}} \right)
\end{align}
where $C_{2k}$ are the Catalan numbers and $D = q^N$ is the dimension of the system's Hilbert space. We refer the reader to Corollary 5 of \cite{brandao2021models} for a technical proof of this bound.

Finally, the probability $\mathrm{Pr}[\mathcal{C}_{\delta} \leq r]$ of Bob obtaining a substantial bias signal $S$ is given by a union bound over all measurements $M$ that can be prepared by depth-$r$ circuits:
\begin{widetext}
\begin{align}
    \mathrm{Pr}[\mathcal{C}_{\delta} \leq r] &:= \mathrm{Pr}\left[ \max_{M} S \geq 1- \frac{1}{D^2} -\delta \right] \nonumber \\
    &= \mathrm{Pr}\left[ \max_{M} \magn{\tr{\left[ M \left( \mathcal{U} \otimes \mathcal{I} - \mathcal{D} \otimes \mathcal{I} \right) \left( \ket{\Psi}\bra{\Psi} \right) \right]} } \geq 1 - \frac{1}{D^2} - \delta \right] \nonumber \\
    &\leq \frac{\magn{M}}{(1-1/D^2-\delta)^{2k}} \mathbb{E} \left[ \magn{\tr{\left[ M \left( \mathcal{U} \otimes \mathcal{I} - \mathcal{D} \otimes \mathcal{I} \right) \left( \ket{\Psi}\bra{\Psi} \right) \right]} }^{2k} \right] \nonumber \\
    &\leq 3 \left(C_{2k} + \frac{\epsilon}{(2k)! D^{3k}} \right) D^2 N^{2r} \magn{G}^r \left( \frac{64 k^4}{D(1-\delta)^2} \right)^k
\end{align}
\end{widetext}
where we assume $1-\delta \geq 2 / D$. In the final line we estimated the number $\magn{M}$ of unique projective measurements that can be performed using depth-$r$ circuits, which is crudely bounded from above by
\begin{equation}
    \magn{M} \leq (2D^2+1) N^{2r} \magn{G}^r
\end{equation}
if these circuits are constructed from a 1- and 2-body gate set $G$ of cardinality $\magn{G}$.

Assuming $N \geq \magn{G}$ and $k \leq D/2$, the probability $\mathrm{Pr}[\mathcal{C}_{\delta} \leq r]$ that Bob obtains a substantial bias signal remains exponentially small until
\begin{equation}
    r \gtrsim \frac{k(N - 4 \log k)}{\log N}.
\end{equation}
We therefore conclude that every $\epsilon$-approximate $k$-design $\mathcal{U}$ has strong circuit complexity $r$ that grows linearly with $k$. Moreover, our calculation of the frame potential Eqs. \eqref{eq:framepotbspin}, \eqref{eq:framepotbsyk}, \eqref{eq:framepotham} presented in the main text demonstrates that our Brownian circuits form an $\epsilon$-approximate $k$-design after a time $2 \Delta t > 4 k \log D + \log \epsilon^{-1/2}$. We therefore immediately conclude that these Brownian circuits generate $\delta$-strong complexity of depth at least $r \sim k N / \log N$ that grows linearly in time $r \sim t$.

\end{document}